\documentclass[journal]{IEEEtran}
\usepackage{amsmath}
\usepackage{bbold}
\usepackage[T1]{fontenc}
\usepackage{amsfonts}
\usepackage{color}
\usepackage{array}
\usepackage{comment}
\usepackage{stfloats}
\usepackage{amsthm}
\usepackage{amssymb}
\usepackage{float}
\usepackage{nccmath}
\usepackage{graphicx}
\usepackage[linesnumbered,ruled]{algorithm2e}
\usepackage{setspace}
\usepackage{booktabs}
\usepackage{subcaption}
\usepackage{enumitem}
\usepackage{mwe}
\usepackage{cite}
\usepackage{amsmath,amssymb,amsfonts}
\usepackage{graphicx}
\usepackage{textcomp}
\usepackage{mathtools}
\usepackage{xcolor}
\usepackage{algorithmicx}
\usepackage{xspace}
\usepackage{lettrine}
\DeclareMathOperator*{\minimize}{minimize}

\DeclareMathOperator*{\wrt}{w.r.t.}

\begin{document}
	\title{Quality-Aware Deep Reinforcement Learning for Streaming in Infrastructure-Assisted Connected Vehicles}
	\author{Won Joon Yun,~\IEEEmembership{Student Member,~IEEE}, Dohyun Kwon, Minseok Choi,~\IEEEmembership{Member,~IEEE},
		\\Joongheon Kim,~\IEEEmembership{Senior Member,~IEEE},
		Giuseppe Caire,~\IEEEmembership{Fellow,~IEEE},
		and
		Andreas F. Molisch,~\IEEEmembership{Fellow,~IEEE}
		\thanks{The work was financially supported in part by the National Research Foundation of Korea under Grants (NRF-2020R1G1A1101164 and NRF-2021R1A4A1030775); and also by the National Science Foundation under award 1816699 (for Andreas F. Molisch).}
		\thanks{W.J. Yun and J. Kim are with the Department of Electrical and Computer Engineering, Korea University, Seoul, Republic of Korea, e-mails: ywjoon95@korea.ac.kr, joongheon@korea.ac.kr.}
		\thanks{D. Kwon is with Hyundai-Autoever, Seoul, Republic of Korea.}    
		\thanks{M. Choi is with the Department of Telecommunication Engineering, Jeju National University, Jeju, Republic of Korea, e-mail: ejaqmf@jejunu.ac.kr.}
		\thanks{A.F. Molisch is with the Department of Electrical and Computer Engineering, University of Southern California, Los Angeles, CA, USA, e-mail: molisch@usc.edu.}
		\thanks{G. Caire is with the Technical University of Berlin, Berlin, Germany, e-mail: caire@tu-berlin.de.}
		\thanks{W.J. Yun and D. Kwon contributed equally to this work (first authors).}
		\thanks{M. Choi is a corresponding author of this paper.}
	}
	\maketitle
	\begin{abstract}
		This paper proposes a deep reinforcement learning-based video streaming scheme for mobility-aware vehicular networks, e.g., vehicles on the highway. 
		We consider infrastructure-assisted and mmWave-based scenarios in which the macro base station (MBS) cannot directly provide the streaming service to vehicles due to the short range of mmWave beams so that small mmWave base stations (mBSs) along the road deliver the desired videos to users. 
	    For a smoother streaming service, the MBS proactively pushes video chunks to mBSs. This is done to support vehicles that are currently covered and/or will be by each mBS. 
		We formulate the dynamic video delivery scheme that adaptively determines 1) which content, 2) what quality and 3) how many chunks to be proactively delivered from the MBS to mBSs using Markov decision process (MDP). 
		Since it is difficult for the MBS to track all the channel conditions and the network states have extensive dimensions, we adopt the deep deterministic policy gradient (DDPG) algorithm for the DRL-based video delivery scheme.
		This paper finally shows that the DRL agent learns a streaming policy that pursues high average quality while limiting packet drops, avoiding playback stalls, reducing quality fluctuations and saving backhaul usage.
	\end{abstract}
	
	\IEEEpeerreviewmaketitle
	
	\section{Introduction}\label{sec:1}
\IEEEPARstart{A}{ccording} to a Cisco report published in 2019, the volume of global mobile data traffic is foreseen to grow to 77 exabytes per month by 2020. 
In addition, 79 percentages of the mobile data traffic are expected to come from mobile video services~\cite{Cisco}. 
In order to tackle such explosive growth of video traffic volume in wireless networks, a large amount of research has been devoted to multiple application scenarios~\cite{XuWC2017, BethTWC2016, HoTMC2017}. 
Among many solutions, video streaming has been considered to be appropriate in delivering such massive video traffic in fifth generation (5G) networks, because streaming does not require the full video to be downloaded at once~\cite{DragoICNC2018}.

Millimeter wave (mmWave) communications in 5G networks are expected to play a key role for high-bit-rate video services thanks to the fact that mmWave systems can achieve very large per-link data rates~\cite{Qualcomm,TVT2021Jung,WC2018Sherman}. 
However, some features of mobile users in vehicular networks, e.g., short connection duration and frequent handoff, make video streaming services still very challenging.
Specifically, in highway scenarios of mmWave vehicular networks, because a single macro base station (MBS) does not have sufficient coverage area, multiple small mmWave base stations (mBSs) by the roadside is required.
In addition, it is difficult for the MBS to directly provide quality-adaptive and seamless video streaming due to time-varying network environments and high mobility of vehicles.

In this paper, we propose a deep reinforcement learning (DRL)-based adaptive video streaming system that employs infrastructures on the road, e.g., small mBSs. 
The proposed streaming system learns a policy of proactively pushing the desired video chunks from the MBS to mBSs before users actually request them, depending on time-varying network states.
We demonstrate by simulation that the learned policy is able to provide smooth and high-quality video streaming to vehicular users even with massive number of vehicles and without any channel information.

\subsection{Motivation}\label{subsec_I_I}
In mmWave-based vehicular networks, the MBS, which has the whole file library, has difficulty to directly provide video contents to all users due to the limited range of mmWave communications.
Also, high mobility of vehicles prevents the MBS from efficiently tracking their channel gains; therefore, roadside-infrastructure-to-vehicle (I2V) video delivery is presented in \cite{TVT2016Kim}. 
Motivated by this system, we consider the scenario in which small mBSs located on the road take a role of relaying the desired contents from the MBS to vehicles in their own coverage regions.
In this infrastructure-assisted vehicular network, high mobility of vehicles requires frequent handoffs; therefore, we focus on proactively pushing video chunks from the MBS to each mBS for vehicles that are in, and/or will soon come into the coverage of the target mBS. 
More specifically, our target scenario is that the MBS pushes upcoming video chunks to mBSs. Vehicle users will play video chunk shortly. When users are already playing the stream, the MBS knows which contents users are playing. 
Since a video stream consists of sequential chunks, the next upcoming chunk to be played should arrive at the queue of the streaming user before its playtime. 
Therefore, our pushing strategy depends on users' playback states and vehicle locations, not on the content popularity profile, which is the main difference from the existing studies on proactive content caching.

Considering the large number of mBSs and frequent handoffs of vehicles, the MBS requires massive network information, e.g., association between vehicles and mBSs and queue states of mBSs, to determine how to push contents to mBSs.
In addition, quality adaptation of video streaming is required for supporting differentiated quality requirements \cite{TON2016Kim}; therefore, the MBS should make decisions also on the quality of videos that will be pushed to mBSs. 
Consequently, there are massive numbers of network states and actions that the MBS has to observe and take, and they increase even more as the number of vehicles grows.
Accordingly, we employ the DRL algorithm that handles multitudinous network states for supporting smooth video streaming services to mobile users.

\subsection{Main Contribution}\label{subsec_I_II}

The main contributions of this paper are as follows:
\begin{itemize}
	\item This paper presents an infrastructure-assisted mmWave vehicular network for smooth and high-quality adaptive video streaming.
	Different from existing studies on content delivery in vehicular networks with the help of entities having own storage, e.g., caching helpers, we consider the scenario in which infrastructures, e.g., mBSs or roadside units (RSUs), support streaming users by relaying video chunks from the MBS having the whole file library. 
	When there are massive numbers of vehicles and the coverage of the mmWave MBS is limited, the MBS has difficulties in directly delivering contents to vehicles. 
	In contrast with existing studies on proactive content caching, pushing contents from the MBS to infrastructures should be instantly adapted depending on the network status and frequent handoffs of vehicles, and we design the scheme applicable to any delivery method via wireless channels.
	
	\item We formulate the adaptive video streaming problem that dynamically determines 1) which content, 2) what quality and 3) how many chunks will be delivered from the MBS to mBSs as a controlled Markov decision process (MDP).
	The deep deterministic policy gradient (DDPG), which is a {\fontfamily{qcr}\selectfont model-free} and {\fontfamily{qcr}\selectfont off-policy} algorithm, is adopted for solving the problem to handle multitudinous network states and the continuous state space. 
	
	\item The proposed adaptive streaming policy satisfies conflicting requirements for QoS of streaming users, i.e., high average quality, small packet drop rates, seamless playback, and smooth quality variations.
	The reward structure of the proposed DRL-based method is carefully designed for balancing and controlling the tradeoff among conflicting performance metrics mentioned above.
	
	\item Extensive simulation results show that our streaming scheme can provide smooth streaming while pursuing high-quality, avoiding playback stalls, limiting packet drops, reducing excessive quality fluctuations, and reducing backhaul usage.
	In addition, for the sake of validation, we show that our streaming policy is well applicable to a real highway data traffic trace acquired from \cite{OTIS_RealTraceOfTraffic}.
\end{itemize}
 

\subsection{Organization of the Paper}\label{subsec_I_III}
The remainder of this paper is organized as follows: Section~\ref{sec:2} summarizes related work. Section~\ref{sec:3} introduces multiple candidate DRL algorithms, including the DDPG algorithm. 
In Section \ref{sec:4}, the infrastructure-assisted mmWave vehicular network is described, and Section~\ref{sec:5} proposes the DDPG-based streaming system. 
Section~\ref{sec:6} provides implementation details of the proposed streaming system and an extensive discussion of the experimental results with respect to various aspects. 
Lastly, Section~\ref{sec:7} concludes this paper.
	\section{Related Work}\label{sec:2}

Significant research has been dedicated to adaptive video streaming in mobile networks. 
In this section, existing studies on two types of adaptive video streaming systems used for various network categories is provided, namely, i) systems based on classical optimization, and ii) systems based on DRL.

\subsection{Classical Streaming Systems}
\label{subsec_II_I}

Bethanabhotla \textit{et al.}~\cite{BethTWC2016} maximized the network utility related to the long-term time-averaged quality for adaptive video streaming while guaranteeing stability of transmission queuing systems.
In addition, Choi \textit{et al.}~\cite{ChoiJSAC2018, ChoiTWC2019} proposed the dynamic delivery scheme for video streaming especially in the wireless caching network, which maximizes the long-term time-averaged quality while avoiding playback delays and achieving stability of user queuing systems.
With the help of device-to-device (D2D) communications, the distributed and dynamic video delivery scheme is proposed in \cite{ChoiTWC2020}.
Sun \textit{et al.}~\cite{TVT2017Sun} introduced the auction-based channel allocation rule for adaptive streaming that provides the proper video quality to vehicles and allows vehicles to compete for channel access opportunities. 
Miller \textit{et al.}~\cite{TM2015Miller} considered video streaming in dense wireless networks for supporting a high number of unicast streaming sessions by jointly optimizing quality adaptation and scheduling of wireless transmissions.
Finally, Yu \textit{et al.}~\cite{TB2017Yu} considered a bitrate adaptive algorithm for variable bitrate (VBR) streaming, which keeps a good balance among several performance metrics and meets heterogeneous user preferences.

\subsection{Recent DRL-based Streaming Systems}
\label{subsec_II_II}



As mentioned above, classical video streaming systems mostly focused on solving optimization problems to determine how to control the delivery options. 
However, as the number of vehicles sharply increases and channel states become difficult to estimate, only heuristic polynomial-complexity methods could solve the problem and its computational complexity increases massively. 
In addition, for time-varying networks, appropriate decisions on the video delivery should be dynamically made and frequently updated in a short time period; however, solving the optimization problem at every slot is time-consuming. 
Therefore, DRL-based video streaming systems for mobile users have been extensively studied. 
Here, the agent of the DRL-based method can instantly determine the delivery options given the network information after learning the deep neural network representing the streaming policy. 

Guo \textit{et al.}~\cite{TVT2019Guo} applied the DRL approach to dynamic resource optimization for designing a wireless video streaming system. 
In \cite{TVT2019Guo}, the wireless buffer-aware video streaming is proposed without the knowledge of channel state information by jointly optimizing bandwidth allocation and buffer management in order to maximize the effective video streaming time of multiple users. 
Bhattacharyya \textit{et al.}~\cite{Mobihoc2019Bhattacharyya} proposed a DRL-based reconfigurable queueing system for video streaming. 
The authors implemented both model-free and model-based reinforcement learning approaches that are tailored to the problem of determining assignments of users to queues in each decision period.
However, \cite{TVT2019Guo} and \cite{Mobihoc2019Bhattacharyya} do not consider dynamic quality selection for receiving chunks so that the average video quality and excessive quality variations are not included in the reward structure either. 

In addition, time-varying radio frequency environments and random service demands are further challenging issues for optimizing the video streaming system. 
To address such challenges, Tang \textit{et al.} \cite{arXiv2019Tang} employed the asynchronous advantage actor-critic (A3C) for proposing a novel two-level cross-layer optimization framework. 
Here, the proposed framework was used at the application layer so that the highly complex multi-user, cross-layer, time-varying video streaming problem can be decomposed into subproblems that can be solved effectively. 
Mao \textit{et al.}~\cite{SIGCOMM2017Mao} developed Pensieve, a system that carries out adaptive flow assignments using reinforcement learning. 
Pensieve can learn the adaptive bitrate (ABR) streaming policy that adapts to a wide range of environmental conditions and quality of experience (QoE) metrics based solely on observations of the resulting performance of past decisions.
However, \cite{TVT2019Guo,Mobihoc2019Bhattacharyya,arXiv2019Tang,SIGCOMM2017Mao} do not consider the vehicular networks in which users have high mobility and the channel information is difficult to be observed in real-time.

Recently, deep neural networks (DNNs) have been used for jointly optimizing quality adaptation and resource allocation for dynamic streaming in wireless networks as shown in \cite{Mobihoc2019Bhattacharyya,TCOMM2019Ye}. 
In \cite{Mobihoc2019Bhattacharyya}, \textit{QFlow}, which is a reinforcement learning approach of selecting bitrates for wireless streaming by adaptively controlling flow assignments to queues, is introduced. 
Power-efficient wireless ABR streaming is proposed in \cite{TCOMM2019Ye} in which power control is jointly optimized with minimization of video transmission time by using deep reinforcement learning (DRL).
In addition, the authors of \cite{TVT2020Guo} proposed the DRL method for quality adaptation and transcoding that balances the tradeoff between the QoE of video services and computational costs of transcoding. 
Similarly, a joint framework of quality adaptation and transcoding is presented in \cite{IoTJ2021Fu} using soft actor-critic DRL, which further reduces bitrate variance.
However, the above studies of \cite{Mobihoc2019Bhattacharyya,TCOMM2019Ye,TVT2020Guo,IoTJ2021Fu} focus on the content delivery from base stations to vehicles different from our goal of proactively pushing contents from the MBS to mBSs before mobile users actually request them.

In \cite{TWC2016Qiao}, a mobility-aware vehicular network in which base stations on the highway provide the video streaming service to mobile users is considered, which motivates our infrastructure-assisted mmWave vehicular network. 
The authors of \cite{TWC2016Qiao} proposed a proactive caching scheme that pushes the desired contents from the server to base stations for supporting the adaptive video streaming in mobility-aware and cache-enabled vehicular networks.
Similarly, DRL-based cache replacement techniques by bringing contents from the media server via backhaul links have been proposed in \cite{CL2019Wu,TCCN2020Zhong,TWC2016Qiao}. 
Different from these work, which proactively cache contents during the off-peak time before users request, this paper focuses on short-term caching, where users are already playing the stream and the mBS knows users' desired contents, upcoming video chunks to be played shortly are pushed to mBSs. 
In our scenario, pushed upcoming video chunks are stored in volatile memory of mBSs and they are removed after being provided to users.
Therefore, our content pushing scenario is dependent on vehicle locations and a chunk sequence of a video file, but not on the popularity profile.

	\section{Deep Reinforcement Learning}
\label{sec:3}

In this section, we briefly introduce the DDPG algorithm and its predecessors, i.e., the deep \textit{Q}-network (DQN) and deterministic policy gradient (DPG) method.
Also, features of these DRL methods regarding their prospective for learning are investigated. 

\subsection{Preliminaries}
\label{subsec_III_I}

An MDP can be defined as $M = \left\{\mathbb{S}, \mathbb{A}, P, R\right\}$, where $\mathbb{S}$ denotes the state space, $\mathbb{A}$ represents the set of possible actions, $P$ signifies the transition probability matrix, and $R$ is the reward function~\cite{Sutton_RL}. In MDP, the agent observes the state $s_{t}$ and makes an action $a_{t}$ in discrete time step $t$. Then, the agent gets the time step reward $r_{t+1}$ from its environment and observes the next state $s_{t+1}$. The time step reward $r_{t+1}$ can be defined with a \textit{time step reward function $R$} of the current state and action, i.e., $r_{t+1} = R(s_{t}, a_{t})$ for the action-value function. 
Here, the goal is to learn the agent's policy in a way that maximizes the expected sum of discounted future reward~\cite{DDPG},
as given by 
\begin{equation}
\label{return}
G_t = \sum_{i=t}^{T}\delta^{i-t} r_i = \sum_{i=t}^T \delta^{i-t} R(s_i, a_i),
\end{equation}
where $\delta \in [0,1]$ stands for the discount factor. 
The agent's action is defined by a stochastic policy, $\pi_{\theta}$, which maps states to a probability distribution over the actions. 
Note that when the policy is implemented as a deep neural network, $\theta$ indicates the weights of the neural network and the policy $\pi$ is parameterized by $\theta$. 
The agent learns a policy $\pi_{\theta}$ by maximizing the sum of discounted future reward $G_t$ for a finite time horizon and we can measure the policy $\pi_{\theta}$ \cite{DDPG}. 
The function $V_{\pi_{\theta}}(s_t)$ quantifies the value of the state at time step $t$ and it can be defined as:
\begin{equation}
\label{V}
V_{\pi_{\theta}}(s_t) = \mathbb{E}_{\pi_{\theta}} \big[ G_t | s_{t} \big]
= \mathbb{E}_{\pi_{\theta}} \Big[ \sum_{i=t}^{T}\delta^{i-t} R(s_i, a_i) \big| s_t \Big],
\end{equation}
where $\mathbb{E}_{\pi_{\theta}}$ indicates the expectation over the policy $\pi_{\theta}$. 
Then, the value function is updated by
\begin{equation}
\label{V_general}
V_{\pi_{\theta}}(s_t) = \mathbb{E}_{\pi_{\theta}}[r_{t+1} + \delta V_{\pi_{\theta}}(s_{t+1})|s_{t}].
\end{equation}

Here, the value of each action is quantified by the so-called \textit{Q}-function, or action-value function. By comparing the value of the \textit{Q}-function for each corresponding action, the agent can estimate how to \textit{behave} on the current state. Compared to the state value function, the \textit{Q}-function considers the action for calculating the value of the action. The \textit{Q}-function is defined as follows:
\begin{equation}
\begin{split}
\label{Q_func}
Q_{\pi_{\theta}}(s_{t}, a_{t}) &= \mathbb{E}_{\pi_{\theta}}[ G_t | s_t, a_t ] \\
&= \mathbb{E}_{\pi_{\theta}}[r_{t+1} + \delta V_{\pi_{\theta}}(s_{t+1})|s_{t}, a_{t}]. 
\end{split}
\end{equation}
Therefore, we can represent the $V_{\pi_{\theta}}(s)$ function as:
\begin{align}
\label{V_withQ}
V_{\pi_{\theta}}(s_t) &= \mathbb{E}_{\pi_{\theta}}[ G_t | s_t ] = \sum_{a_t \in \mathbb{A}} \pi_{\theta}(a_t|s_t) Q_{\pi_{\theta}}(s_t, a_t),
\end{align}
where $\pi_{\theta}(a_t|s_t)$ is the policy, defined as a probability distribution over the actions given the state.
In order to optimize the policy $\pi_{\theta}$, a policy gradient approach, sets an objective function $\mathcal{J}(\theta)$ with respect to $\theta$, which represents the agent's expected reward in the form: 
\begin{align}
\label{rl_obj}
\mathcal{J}(\theta) &= \sum_{s\in\mathbb{S}}d_{\pi_{\theta}}(s)V_{\pi_{\theta}}(s) \nonumber \\
&= \sum_{s\in\mathbb{S}}d_{\pi_{\theta}}(s)\sum_{a\in\mathbb{A}}\pi_{\theta}(a|s)Q_{\pi_{\theta}}(s, a),
\end{align}
where $d_{\pi_{\theta}}(s)$ stands for the stationary distribution of the controlled Markov chain (MC), i.e., $d_{\pi_{\theta}}(s) = \lim_{t\to\infty}P(s=s_{t}|s_{0}, \pi_{\theta})$, where $P(s=s_{t}|s_{0}, \pi_{\theta})$ is the probability that the the chain visits state $s = s_{t}$ given the initial state $s_{0}$ and the policy $\pi_{\theta}$. 
Then, the adaptive policy $\pi^*_{\theta}$ is obtained as
\begin{equation}
\label{opt_pi}
\pi^*_{\theta} = \arg\max_{\pi_{\theta}} \mathcal{J}(\theta).
\end{equation}
According to the policy gradient theorem \cite{Sutton1999NIPS}, the gradient of the objective function $\mathcal{J}(\theta)$ is given by
\begin{equation}
\begin{split}
\label{rl_obj_grad}
\nabla_{\theta}\mathcal{J}(\theta) 
&= \nabla_{\theta}\sum_{s\in\mathbb{S}}d_{\pi_{\theta}}(s)\sum_{a\in\mathbb{A}}Q_{\pi_{\theta}}(s, a)\pi_{\theta}(a|s)\\
&\propto \sum_{s\in\mathbb{S}}d_{\pi_{\theta}}(s)\sum_{a\in\mathbb{A}}Q_{\pi_{\theta}}(s, a)\nabla_{\theta}\pi_{\theta}(a|s)\\
&= \sum_{s\in\mathbb{S}}d_{\pi_{\theta}}(s)\sum_{a\in\mathbb{A}}\pi_{\theta}(a|s)Q_{\pi_{\theta}}(s, a)\frac{\nabla_{\theta}\pi_{\theta}(a|s)}{\pi_{\theta}(a|s)}\\
&= \mathbb{E}_{\pi_{\theta}}[Q_{\pi_{\theta}}(s, a)\nabla_{\theta}\ln{\pi_{\theta}(a|s)}],
\end{split}
\end{equation}
and DRL optimizes $\theta$ by utilizing the gradient ascent method.

\subsection{Deep Q-Network (DQN)}
\label{subsec_III_II}

The DQN is one of the breakthroughs of deep reinforcement learning algorithms, which uses the neural network to approximate the \textit{Q}-function and enables it to learn a policy even with the high-dimensional state space.
The agent collects its experiences paired with $(s_{t}, a_{t}, r_{t+1}, s_{t+1})$ and stores them into a replay buffer $\mathcal{D}$. 
Then, the DQN agent iteratively samples a bundle (i.e., a batch) of the experience information $(s_{t}, a_{t}, r_{t+1}, s_{t+1})$ from the replay buffer.
Specifically, with independent samples $(s_{t}, a_{t}, r_{t+1}, s_{t+1})$ for all $t$, the \textit{target} \textit{Q}-function is computed by $\hat{Q}(s_{t}, a_{t} ; \theta^{-}) = r_{t+1} + \delta \max_{a_{t+1}}\hat{Q}(s_{t+1}, a_{t+1} ; \theta^{-})$, where $\theta^-$ indicates parameters of the target network. 
Then we update $\theta$ by using the gradient of 
\begin{equation}
L(\theta) = (\hat{Q}(s_{t}, a_{t}; \theta^{-}) - Q(s_t, a_t ; \theta))^2.
\end{equation}
Here, $\theta$ is repeatedly updated for $N_{\text{target}}$ learning steps (i.e., episodes), on the other hand, $\theta^-$ is updated once every $N_{\text{target}}$ episode by letting $\theta^- = \theta$; therefore, the target network is fixed for $N_{\text{target}}$ episodes. 
However, DQNs suffer from the overestimation issue \cite{arXivPreprintQLearning}, because it always takes the estimate of the maximum Q value. Such overestimated action values could lead to unstable learning and low accuracy.

\subsection{Deterministic Policy Gradient (DPG)}
\label{subsec_III_III}

The DPG~\cite{ICML2014DPG} algorithm attempts to learn a deterministic policy $\mu(s)$ that directly maps the state $s$ of the agent into an action $a$, i.e., $a = \mu(s)$. 
When the state space is very large, learning a deterministic policy is easier than learning a stochastic policy. 
According to \cite{ICML2014DPG}, the DPG algorithm optimizes the $\mu_{\theta}$ that is parameterized with neural network weights $\theta$ by using the gradient as follows:
\begin{equation}
\label{DPG_grad}
\nabla_{\theta}\mathcal{J}(\theta) = \sum_{s \in \mathbb{S}} d_{\mu_{\theta}} (s) \nabla_{\theta} \mu_{\theta}(s) \nabla_a Q_{\theta}(s,a)|_{a=\mu_{\theta}(s)}.
\end{equation}
By using the policy gradient approach, the DPG updates the $\theta$ to optimize the deterministic policy of $\mu_{\theta}$.

\subsection{Deep Deterministic Policy Gradient (DDPG)}
\label{subsec_III_IV}

The DDPG is an {\fontfamily{qcr}\selectfont off-policy} \textit{actor-critic} algorithm that combines advantages of the DQN and the DPG. 
The off-policy stands for a policy that has two independent \textit{Q}-functions as the DQN. 
The \textit{actor} holds a policy $\mu_{\theta}$ and is responsible for the agent's action. 
On the other hand, the \textit{critic} assesses the action derived from the \textit{actor}. 
By interacting with each other, the \textit{critic} learns the \textit{Q}-function to accurately evaluate the action of the \textit{actor}, while the \textit{actor} learns the policy from the \textit{critic}'s \textit{Q}-function. To optimize both \emph{actor} and \emph{critic}, the target network is devised, which consists of \emph{target actor} and \emph{target critic}.

The agent's experience $(s_{t}, a_{t}, r_{t+1}, s_{t+1})$ sampled from the replay buffer $\mathcal{D}$, is utilized to update the parameter $\theta^{{Q}}$ of the \textit{critic}.             
The experiences are sampled with size of $N$ and are exploited to calculate the \textit{critic}'s loss function as follows:
\begin{align}
\label{Loss_ddpg_critic}
\minimize~ &L(\theta^{Q}) = \frac{1}{N}\sum^N_{i}(y_i - Q(s_{t}, a_{t}|\theta^{Q}))^2,\\
\label{Loss_ddpg_constraint}
\wrt~ &\space y_i = r_{t+1} + \delta \hat{Q}(s_{t+1}, \mu(s_{t+1}|{\theta^{\hat{\mu}}})|\theta^{\hat{Q}}),
\end{align}
where, $\theta^{\hat{\mu}}$ stands for the weights of the target actor network. The parameter $\theta^{\mu}$ of \emph{actor} is updated using the following sampled policy gradient:
\begin{align}
\label{DDPG_actorupdate}
&\nabla_{\theta^\mu}\mathcal{J}(\theta) \nonumber \\ 
&~~\approx \frac{1}{N}\sum_{i}{\nabla_{a}Q(s, a|\theta^{Q})}|_{s=s_i, a=\mu_{\theta}(s_i)}\nabla_{\theta^{\mu}}\mu(s|\theta^{\mu})|_{s_{i}},
\end{align}
where $\nabla_{a}Q(s, a|\theta^Q)|_{s=s_i, a=\mu(s_i)}$ is input of the \emph{critic} network. 
Finally, the target network parameters (\emph{i.e.}, $\theta^{\hat{Q}}$ and $\theta^{\hat{\mu}}$) are \textit{soft} updated, where $0 < \tau \ll 1$ as follows:
\begin{equation}
\begin{split}
\label{target_update}
\theta^{\hat{Q}} &\leftarrow \tau\theta^{Q} + (1-\tau)\theta^{\hat{Q}},\\
\theta^{\hat{\mu}} &\leftarrow \tau\theta^{\mu} + (1-\tau)\theta^{\hat{\mu}}.
\end{split}
\end{equation}
In the following sections, the adaptive video streaming policy in mmWave vehicular networks is modeled by the MDP. In order to achieve efficient and improved learning of the streaming policy, we adopt a DDPG-based method because the number of states is massive and the state space is continuous in the considered network model.
	\begin{figure}[t]
	\centering
	\includegraphics[width =0.99\linewidth, height =5cm]{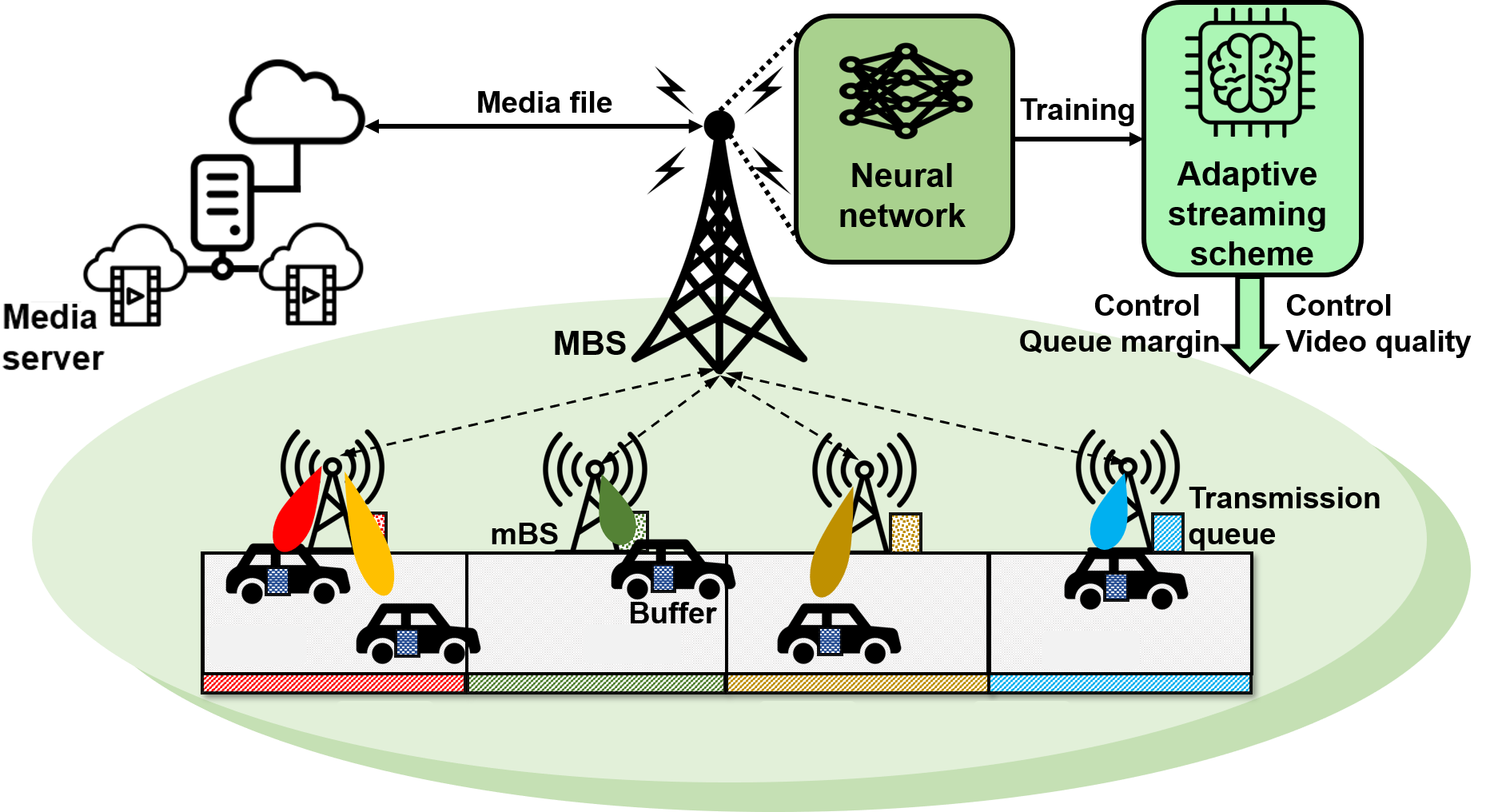}
	\caption{Adaptive streaming system in mmWave vehicular networks on highway.}
	\label{fig:fig1}
	\vspace{-3mm}
\end{figure}

\section{mmWave Vehicular Networks with Adaptive Video Streaming}
\label{sec:4}

In this section, a mmWave-based vehicular networks supporting adaptive video streaming services is introduced.
Consider a mobility-aware scenario, e.g., highway, where there is bidirectional road traffic and $\mathcal{N}$ vehicles are moving in each direction.
Although bidirectional traffic is considered for simulation, we will describe the vehicular network and explain our proposed streaming policy for the traffic with one direction.
However, the problem formulation and the proposed streaming policy can be easily extended to cover the bidirectional traffic by scaling the state and action spaces.
$\mathcal{K}$ mBSs are deployed at regular intervals along with the highway, as shown in Fig. \ref{fig:fig1}.
Denote the sets of vehicles and mBSs as $\mathcal{U} = \{u_{1} \cdots, u_{\mathcal{N}}\}$, and $\mathcal{X} = \{ x_{1}, \cdots, x_{\mathcal{K}}\}$, separately, where $u_i$ and $x_j$ represent the $i$-th vehicle and $j$-th mBS.
Each mBS has its own coverage region and can communicate with vehicles only in its coverage region.
Suppose that coverage regions of mBSs are exclusive and 
every $u_i$ can be associated with only one mBS at each time slot. Here, denote the set of vehicles in coverage of $x_j$ as $\mathcal{U}_j$.

Considering that 
line-of-sight (LOS) between an mBS and a vehicle is available in most of the highways, mBSs can directly conduct beam alignment towards vehicles in their own coverage regions.
Suppose that mmWave beams are highly directive and multiple vehicles associated with the same mBS are sufficiently separated so that interference among mmWave beams are negligible even in the identical coverage region~\cite{ton2011singh}. 
In addition, all mBSs are connected to a central MBS, and the MBS has an access to media server, which stores all contents. 
The MBS can gather all the network information from mBSs via backhaul connections. 
Similarly, the MBS is capable of pushing the desired contents to mBSs via backhaul links with significant latency; therefore, instantaneous streaming from the MBS to the vehicles via the mBS is not possible. 

In this network, we focus on the scenario in which vehicles on the highway are requesting dynamic video streaming services. 
A stream consists of sequential chunks and each chunk is in charge of the fixed playtime.
Also, dynamic streaming allows users to play the video before receiving the whole stream; therefore, users continuously request the desired chunks while playing the previously received chunks. 
We assume that the MBS having the whole file library cannot transmit the video directly to vehicles because the range of mmWave communications is limited. 
In addition, if there are large numbers of vehicles with high mobility on the highway, it is very difficult for the MBS to directly communicate with all vehicles and to track its beams to all vehicles.
On the other hand, mBSs on the highway can be considered as small RSUs which do not have any access to the whole file library. 
In summary, the MBS proactively delivers the desired video chunks to mBSs via backhaul links, and each mBS provides chunks with appropriate quality to vehicles in its coverage region depending on the network state. 
Here, proactively pushed video chunks from the MBs are accumulated in transmitter queues at mBSs, and transmitter queue states of mBSs are updated at every time slot by receiving chunks from the MBS as well as delivering them to their associated vehicles.

In this context, there are three main decisions to be made for supporting smooth and high-quality video streaming services in the vehicular network: 1) which content, 2) what quality, and 3) how many chunks will be pushed from the MBS to each mBS. 
Note that the content delivery from mBSs to vehicles is not considered due to the following reasons. 
First, the content delivery from mBSs to vehicles should be real-time; however, gathering all network states and decision-making at the MBS are time-consuming. 
Meanwhile, proactive pushing of contents from the MBS to vehicles does not have to be in the same timescale of real-time streaming. Second, if the MBS can jointly determine how
to deliver video chunks from the MBS to mBSs as well as from mBSs to vehicles, then more exact information is required and more decisions should be carefully made.
For example, exact positions and velocities of vehicles and channel estimation are required for determining the quality and amounts of chunks to be delivered to vehicles.
In addition, resource allocation and/or scheduling should be designed with the delivery of chunks.
Accordingly, more states and actions make the decision-making process much more time-consuming.

Since smooth playback is one of the most important factors determining quality-of-service (QoS) in streaming systems and the backhaul usage is expensive, having the MBS to serve all requests (although routed through the mBS) in a purely reactive way could be inefficient and subject to congestions in the backhaul. 
Therefore, our proposed adaptive video streaming system learns how to proactively push the desired contents, i.e., which contents, what quality and how many chunks are required 
for the following purposes: 1) pursuing high-quality streaming, 2) avoiding playback stalls, 3) limiting packet drop rates, and 4) reducing extreme quality fluctuations; however, these purposes are conflicting so that we have to design the reward function carefully and this will be explained in detail in the following sections.
The detailed system descriptions and assumptions on elements in this network follow.
\begin{figure}[t]
	\centering
	\includegraphics[width=0.905\columnwidth]{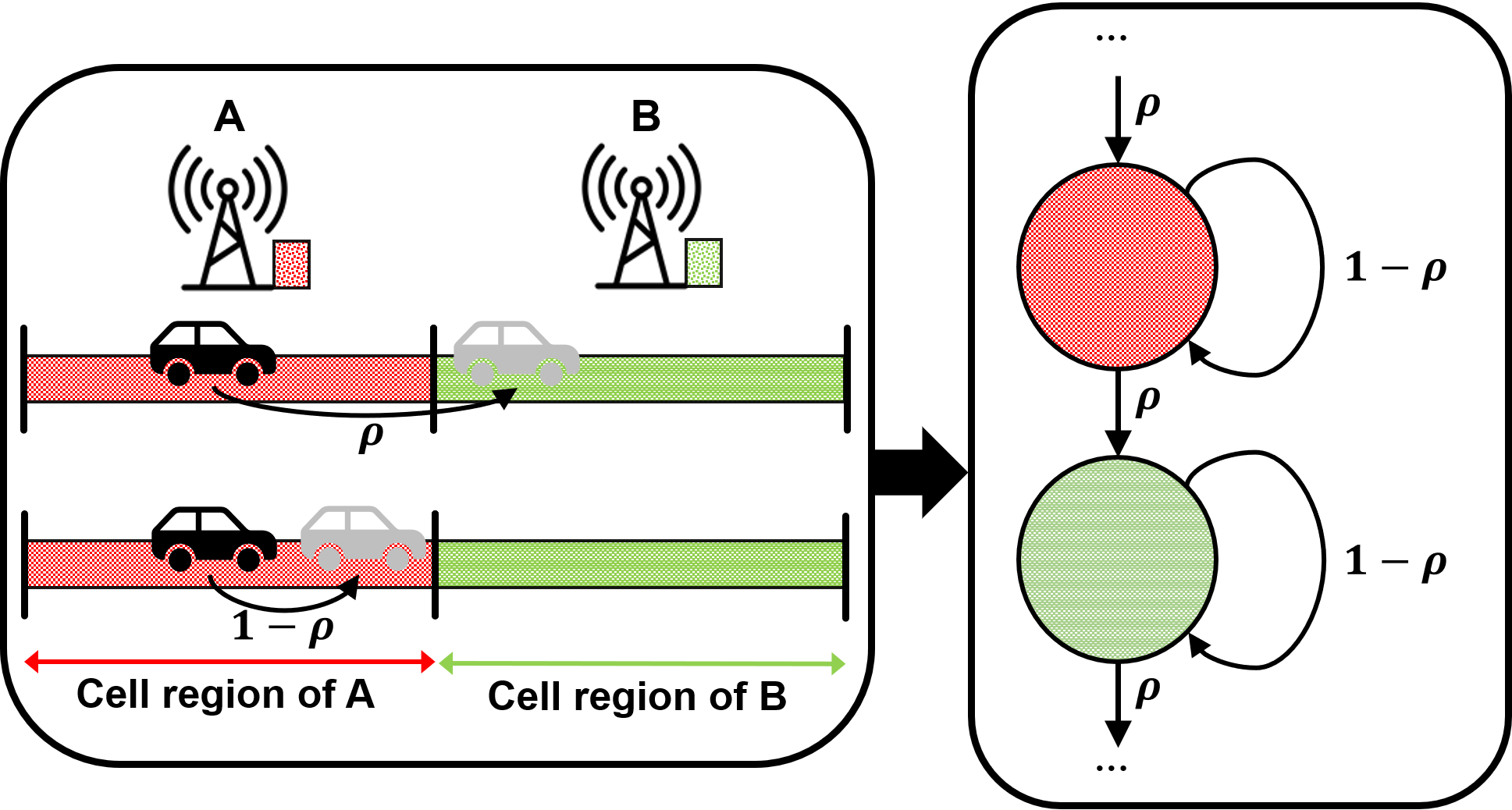}
	\caption{Model of the finite state Markov chain (FSMC) for vehicle position transition. Note that all vehicles have the same FSMC model with respect to system parameters such as the diameter of the cell coverage region and velocity of each vehicle.}
	\label{fig:fig2}
\end{figure}

\subsection{System Description}
\label{subsec_IV_I}

\subsubsection{\textbf{Vehicles on highway}}
On the highway, each vehicle $u_{i}$ can only move forward, i.e., if $u_{i}$ is in the coverage of mBS $x_{j}$ at time $t$, $u_{i}$ cannot associate with any $x_{n}$ for all $n\in\{1,\cdots,j-1 \}$ at time $t' > t$.
Denote the mBS index that vehicle $u_i$ is associated with at slot $t$ by $p_{i}(t) \in \{0,1,\cdots,\mathcal{K},\mathcal{K}+1 \}$.
$p_i(t) = 0$ indicates that $u_i$ does not appear on the highway yet, and $p_i(t) = \mathcal{K}+1$ means that $u_i$ has left this highway.

\begin{figure}[t]
	\centering
	\includegraphics[width =0.99\linewidth, height = 7.2cm]
	{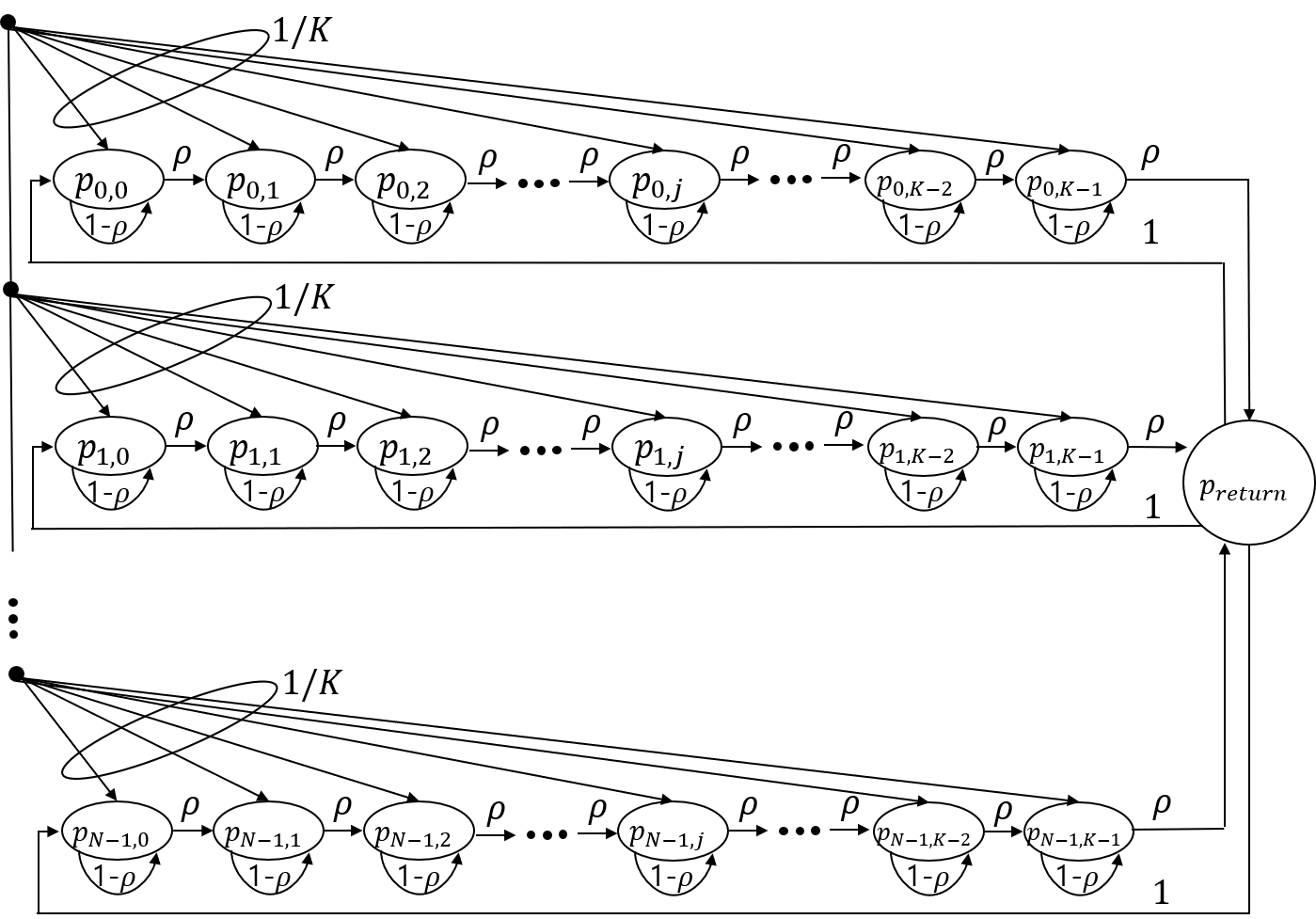}
	\caption{Finite state Markov chain (FSMC) of position transition model of vehicles on considered mmWave vehicular networks.}
	\label{fig:MarkovChain}
	\vspace{-3mm}
\end{figure}

The time-varying vehicle positions, i.e., $p_{i}(t)$ for all $i \in \{1, \cdots, \mathcal{N}\}$, are modeled by the finite state Markov chain (FSMC) depicted in Fig.~\ref{fig:fig2}. 
In the FSMC model, a vehicle associated with $x_j$ can move into the coverage region of the next mBS $x_{j+1}$ with the transition probability $\rho$ that can be derived from the system average velocity $v$ and the coverage range $\mathcal{O}$ of each mBS.
On the other hand, a vehicle can stay in the same region of $x_j$ with probability of $1-\rho$; therefore, if mobility is very high and the coverage region of each mBS is small, then $\rho$ would be close to one.
Note that the traffic model can be illustrated with Markov chain in Fig.~\ref{fig:MarkovChain}. 
Since the unit of $v$ is km/h, the transition probability is assumed to be $\rho=\frac{1000v}{3600 \mathcal{O}}$.
Suppose that all vehicles are moving with at average velocity $v=80$ km/h, and the range of each mBS cell is $\mathcal{O} = 150$ m, then $\rho = 0.143$ in this paper.

\begin{table}[t]
	\caption{System description parameters}
	\footnotesize
	\label{tab:tab1}
	\begin{center}
		\centering
		\begin{tabular}{c||l}
			\toprule[1.0pt]\centering
			Parameter & Description\\
			\midrule
			\midrule
			$\mathcal{N}$ & The number of vehicles\\
			$\mathcal{K}$ & The number of mBSs\\
			$\mathcal{M}$ & The macro base station \\
			$\mathcal{X}$ & The set of mBSs \\
			$\mathcal{U}$ & The set of vehicles \\    
			$x_j$ & The $j$-th mBS \\
			$u_i$ & The $i$-th vehicle \\
			$p_i$ & mBS association of $u_i$ \\
			$c_{j}^{(i,q)}$ & Queue length of $x_j$ for $u_i$ and quality $q$ \\
			$l_{j}^{(i,q)}$ & chunks with quality $q$ delivered from MBS to $x_j$ for $u_i$ \\
			$w_{j}^{(i,q)}$ & chunks with quality $q$ delivered from $x_j$ to $u_i$ \\
			$b_i$ & Buffer length of $u_i$ \\
			$\mathcal{F}$ & Playback rate of vehicular buffer \\
			$h_i$ & Sum of delivered chunks for $u_i$ \\
			$\bar{q}_i$ & Average quality of arrived chunks at $u_i$ \\
			$\mathcal{Q}$ & The set of video quality \\  
			$L$ & Number of possible quality levels \\
			$\overline{c}$ & The finite size of mBS queue\\     
			$\overline{b}$ & The finite size of the vehicular buffer \\
			$\rho$ & The FSMC transition probability\\
			\midrule
			$\gamma$ & The learning rate of $\mathcal{M}$ \\
			$s_t$ & State at $t$ in MDP \\
			$a_t$ & Action at $t$ in MDP \\
			$r_t$ & Reward at $t$ in MDP \\
			\bottomrule[1.0pt]
		\end{tabular}
	\end{center}
\end{table}

\subsubsection{\textbf{Transmitter Queue at mBS}}
Each mBS $x_j$ is equipped with a video transmission queue. 
Denote $c_j^{(i,q)}(t)$ as the transmission queue length of $x_j$ at slot $t$ that accumulates the chunks with quality $q$ requested by $u_i$. 
The finite queue size, denoted by $\overline{c}$, is identical for all mBSs. 
The number of chunks transmitted to vehicles are considered as departures of the queue, and it obviously depends on data rates of links with vehicles. 
Then, the transmitter queue of $x_j$ evolves at slot $t$ for all $u_i \in \mathcal{U}_j$ and $q\in \mathcal{Q}$ as follows:
\begin{align}
&c_j^{(i,q)}(t+1) = \nonumber \\
&~\min\big\{ \max\{ c_j^{(i,q)}(t) - w_j^{(i,q)}(t), 0 \} + l_j^{(i,q)}(t), \bar{c} \big\}, 
\end{align}
where $w_{j}^{(i,q)}(t)$ and $l_{j}^{(i,q)}(t)$ represent numbers of chunks with quality $q$ that mBS $x_j$ delivers vehicle $u_i$ and that the MBS pushes to $x_j$ for $u_i$.
Note that $c_j^{(i,q)}(t) \leq \bar{c}$ for all $x_j \in \mathcal{X}$, $u_i \in \mathcal{U}$, $q\in \mathcal{Q}$ and $t\in\{1,\cdots,T \}$.

If too many chunks are delivered from the MBS to mBS $x_j$ and the queue cannot take all of the delivered chunks, i.e., $\max\{ c_j^{(i,q)}(t) - w_j^{(i,q)}(t), 0 \} + l_j^{(i,q)}(t) > \bar{c}$, we call this event as \textit{packet drop} at mBS. 
The packet drop event at mBS is consistent with a waste of backhaul, and the proposed adaptive video streaming will learn the policy to limit the packet drop rate.
Here, if there exists a vehicle $u_i$ that moves forward to the coverage of the next mBS in slot $t+1$, the remaining chunks in its queue $c_j^{(i,q)}(t+1)$ for all $q\in \mathcal{Q}$ are assumed to be discarded. 
In fact, since the mBS determines the action depending on the information of vehicles that are in its coverage and will come in its coverage soon, 
the remaining chunks for vehicles that moved forward to the next mBSs earlier do not affect the current action. 
Instead, the discarded chunks can be considered as wastes of backhaul usage because chunks pushed from the MBS become useless.

\subsubsection{\textbf{Receiver buffer at vehicle}}
Each vehicle $u_i$ is equipped with a receiver buffer in which arrived video chunks are waiting to be played. 
Denote the buffer length of $u_i$ at slot $t$ by $b_i(t)$ for all $u_i \in \mathcal{U}$, and buffer lengths of all vehicles are assumed to be identical, denoted by $\bar{b}$. 
Since the playback rate of video streaming does not generally change, we assume that the number of chunks played in each time slot is a constant $\mathcal{F}$.
Then, the buffer dynamics of $u_i$ can be represented as follows:
\begin{equation}
b_i(t+1) = \max\{ b_i(t) - \mathcal{F}, 0\} + \sum_{q\in\mathcal{Q}} w_{p_i(t)}^{(i,q)}(t).
\end{equation}
Note that $w_{p_i(t)}^{(i,q)}(t)$ is the number of chunks with quality $q$ delivered from the currently associated mBS $x_{p_i(t)}$ to vehicle $u_i$, and each chunk can have different qualities in the dynamic streaming service.
Owing to a constant $\mathcal{F}$, a buffer can be empty if arrivals are not sufficiently large.
If there is no chunk to be played in the buffer, i.e., $b_i(t) = 0$, the user experiences video \textit{playback stall}.
Similar to packet drop at mBS, if too many chunks arrive and the buffer size is not sufficient to store all these chunks, i.e., $b_i(t) \geq \bar{b}$, some of them have to be dropped, and we would call this event \textit{packet drop at vehicles}.
However, chunk delivery from mBSs to vehicles is not directly determined by the learning agent (i.e., MBS).
Accordingly, packet drop at vehicles will not be included in the reward structure of the DDPG algorithm, but the simulation results in Sec \ref{sec:6} will show that the proposed video streaming can also reduce the packet drop events at vehicles.

\subsubsection{\textbf{Video content}}
Dynamic video streaming is considered, and each chunk can have a different quality level. 
Suppose that there are $L$ quality levels, and the set of video qualities is denoted by $\mathcal{Q}=\{q_1, \cdots, q_L\}$.
Here, the bitrate of a video is adopted as a quality measure.
The average link capacity, which is required for transmitting a chunk with quality level $q_l$ is denoted by $\tilde{C}_l$.
In addition, we assume that each chunk plays $\tau$ seconds of the entire stream; therefore, the bitrate $q_l$ is equal to $\tilde{C}_l / \tau$.

	\textit{Remark}: In general, uploading all the network states to the MBS via backhaul links could occur with significant latency; however, note that the proposed streaming scheme works well with the latency of backhaul communications because of the following two reasons. 
	First, the content delivery from the MBS to mBSs, which is the action of the proposed DDPG agent is basically for supporting the near future streaming of vehicle users who are in, or will soon move into the coverage of a target mBS, not for the real-time streaming. 
	In other words, the proposed DDPG agent determines to proactively push the desired video chunks from the MBS to mBSs before the user actually requests them; therefore, some latency can be acceptable.
	
	Second, although the video streaming is real-time, the content delivery from the MBS to mBSs does not have to follow this timescale, and the MBS does not have to deliver the desired chunks to mBSs very frequently. 
	In this case, the buffer states of vehicles that the MBS receives could be outdated, but this can be easily adapted because the playback rate is generally not changed. 
	For example, let each chunk play $\tau$ milliseconds in the whole stream and uploading the network information to the MBS via backhaul links takes at least $a \cdot \tau$ milliseconds, where $a$ is a positive integer value. 
	However, this can be easily handled by considering the timescale of pushing chunks from the MBS to mBSs is $a$ times of the timescale of delivering chunks from mBSs to vehicles.  
	Then, the DDPG agent naturally learns the streaming policy to accumulate more than $a$ chunks in vehicle buffers.
	Accordingly, the long latency of backhaul links does not spoil the necessity of the proposed dynamic streaming scheme and also the proposed scheme is tolerable to some degree of backhaul latency.

\subsection{Assumptions}\label{subsec_IV_II}
In this subsection, some necessary assumptions are introduced for the following components: MBS, mBSs, vehicles, and video contents. Note that such assumptions are not particularly restrictive and are generally satisfied in practice.

\begin{itemize}
	\item \textbf{MBS}: 
	The MBS has and access to the whole file library and can send any content to mBSs via backhaul links. 
	We assume that the backhaul has the infinite capacity, but in fact, the transmitter queue length of each mBS is limited to $\bar{c}$ chunks; therefore, the maximum number of chunks that can be delivered from the MBS to a mBS is $N \cdot L\cdot \bar{c}$.
	In the proposed mmWave vehicular network, the MBS takes the role of a learning agent for the adaptive video streaming system for vehicles on the highway. 
	Suppose that the MBS gathers the information of the considered vehicular network from mBSs in the following four respects: 1) the association between mBSs and vehicles, 2) the transmission queue state of each mBS, 3) the buffer state of each vehicle, and 4) the average quality of the video chunks previously delivered to vehicles. 
	Exploitation of this information enables the MBS to learn the adaptive streaming policy through considerable trial and error for the observed state of the vehicular network. 
	The streaming option consists of the following decisions: 1) which content, 2) what quality and 3) how many chunks will be pushed from the MBS to mBSs. 
	
	\item \textbf{mBSs}: 
	The mBS is capable of proactively accumulating video chunks in its queues, requested by vehicles in its coverage region as well as those that will be expected to arrive in its coverage region shortly.
	Here, when the MBS updates the queue state of mBS $x_j$, we simply assume that $x_j$ is willing to request the video chunks from the MBS for the vehicles in its coverage region and the region of the mBS $x_{j-1}$, i.e., $\mathcal{U}_j$ and $\mathcal{U}_{j-1}$.
	If there are $N_j$ and $N_{j-1}$ vehicles in the coverage of mBSs $x_j$ and $x_{j-1}$, respectively, then queues for $N_j \cdot N_{j-1}$ vehicles at mBS $x_j$ would be updated in the current time slot. 
	Therefore, mBS $x_j$ has $N_j \cdot N_{j-1} \cdot L$ active transmitter queues in every time slot.
	We suppose that video chunks arrived at the mBS should be stored in its transmitter queue for at least one time slot; otherwise, it appears that chunks are directly delivered from the MBS to vehicles. 
	
	When the mBS delivers the desired chunks to vehicles, the `high-quality first' rule is assumed. 
	Depending on the randomly generated channel condition between $x_j$ and $u_i$, the high-quality chunks have a high priority to be delivered to vehicles. 
	For example, consider the queue state of $c_j^{(i,1)}=0$, $c_j^{(i,2)}=0$, $c_j^{(i,3)}=5$, $c_j^{(i,4)}=5$, and $c_j^{(i,5)}=1$. 
	With $\tilde{C} = \{ \tilde{C}_1=1, \tilde{C}_2=3, \tilde{C}_3=5, \tilde{C}_4=8, \tilde{C}_5 = 40 \}$ Mbps, if the channel capacity between $x_j$ and $u_i$ is 61 Mbps, then $w_j^{(i,1)} = 0$, $w_j^{(i,2)} = 0$, $w_j^{(i,3)} = 1$, $w_j^{(i,4)} = 2$, and $w_j^{(i,5)} = 1$ are determined.
	
	\item \textbf{Vehicles}: It can be envisioned that vehicles on the highway only move forward, i.e., a vehicle would transfer to the coverage region of the next mBS or stay within the coverage region of the currently associated mBS after a time step. 
	We assume that vehicles move forward according to the FSMC model, as explained in Section \ref{subsec_IV_I} (see Fig. \ref{fig:fig2}). 
	Vehicles randomly request the video contents, and the MBS knows their desired contents. 
	While playing the stream, vehicles continuously receive the next chunks from their associated mBSs and the received chunks are accumulated in the receiver buffer. 
	
	\item \textbf{Video contents}: 
	Each video content consists of multiple chunks, and each chunk can have different qualities in adaptive video streaming. 
	All chunks during a session are deterministically requested in sequence and they are delivered toward corresponding vehicles. 
	If some chunks fail to be provisioned toward vehicle, they are continuously requested in successive slots. 
	For each video chunk, it is assumed that the quality of a video chunk determines its size, as explained in Section \ref{subsec_IV_I}. 
\end{itemize}

\section{Proposed DDPG-based quality-aware adaptive video streaming system} \label{sec:5}

In this section, the adaptive streaming policy characterized by pushing the desired video chunks from the MBS to mBSs is designed by an MDP, and the DDPG-based DRL algorithm provides decisions on the adaptive streaming policy. 
For simplicity, the learning agent (i.e., MBS) is denoted by $\mathcal{M}$.
This agent adaptively determines 1) which content, 2) what quality, and 3) how many chunks will be delivered from the MBS to each mBS in each time slot.
Based on the random channel conditions, $\mathbf{l}_j^i = [l_j^{(i,1)}, \cdots, l_j^{(i,L)}]$ 
for all $x_j \in \mathcal{K}$ and $u_i \in \mathcal{U}_j$, satisfying $p_i(t) = j$ or $p_i(t) = j-1$, are determined by the DDPG-based DRL algorithm. 
The learning process is introduced by describing the state, the action, random events, state transition, the reward structure, and the algorithm in detail.

\subsubsection{\textbf{State}}
There are four different states of the mmWave vehicular network at time slot $t$ as follows: 1) positions of vehicles (i.e., which mBS is associated with vehicles), 2) transmitter queue states of mBSs, 3) receiver buffer states of vehicles, 4) the total number of delivered chunks to each vehicle until slot $t$, and 5) average qualities of chunks that vehicles received until slot $t$. 
Note that all of the states have a Markov property because
the FSMC model is assumed for mBS associations of vehicles and queue and buffer lengths are dependent on their lengths at the previous slot.
In addition, the average quality state of the already delivered chunks to $u_i$ until slot $t$ depends on its previous value as given by
\begin{equation}
\bar{q}_i(t) = \frac{1}{h_i(t)} \Big( h_i(t-1) \bar{q}_i(t-1) + \sum_{q\in\mathcal{Q}} q\cdot w_{p_i(t)}^{(i,q)}(t) \Big),
\end{equation}
where $h_i(t) = \sum_{\tau = 1}^{t} \sum_{q\in\mathcal{Q}} w_{p_i(t)}^{(i,q)}(\tau)$ indicates the total number of the received chunks until slot $t$; therefore, $\bar{q}_i(t)$ is the average quality per received chunk until $t$. 

Note that the states of $h_i(t)$ and $\bar{q}_i(t)$ are used for reducing excessive quality variations.
Not only qualities of individual chunks but also quality variations during the stream session strongly influence on users' satisfactions. 
Even though the average quality of the received chunks is high, sharp variations of quality levels of successive chunks are detrimental to the user's QoE in video playback. 
Therefore, we intentionally control the video quality to be increasingly improved depending on $h_i(t)$ and $\bar{q}_i(t)$, as the playback session goes by; the details will be explained in the reward part. 

Thus, the state of the proposed MDP at slot $t$ is defined as $s_t \triangleq \{ \mathbf{p}(t), \mathbf{c}(t), \mathbf{b}(t), \mathbf{h}(t), \bar{\mathbf{q}}(t) \}$, where $\mathbf{p}(t) = [p_1(t), \cdots, p_{\mathcal{N}}(t)]$, $\mathbf{c}(t) = [c_j^{(i,q)}(t): u_i\in \mathcal{U}_j \cup \mathcal{U}_{j-1}, x_j \in \mathcal{X}, q\in \mathcal{Q} ]$, $\mathbf{b}(t) = [b_1(t), \cdots, b_{\mathcal{N}}(t)]$, $\mathbf{h}(t) = [h_1(t), \cdots, h_{\mathcal{N}}(t)]$, and $\bar{\mathbf{q}}(t) = [\bar{q}_1(t), \cdots, \bar{q}_{\mathcal{N}}(t)]$.
Here, $\mathbf{p}(t) \in \mathbb{P} \triangleq \{0,1,\cdots,\mathcal{K}, \mathcal{K}+1 \}^{\mathcal{N}}$, $\mathbf{c}(t) \in \mathbb{C} \triangleq \{0,1,\cdots,\bar{c} \}^{\mathcal{K}\cdot \mathcal{N} \cdot L}$, $\mathbf{b}(t) \in \mathbb{B} \triangleq \{0,1,\cdots,\bar{b} \}^{\mathcal{N}}$, $\mathbf{h}(t) \in \mathbb{H} \triangleq \big\{ \{0 \} \cup \mathbb{I}^+ \big\}^{\mathcal{N}}$, and $\bar{\mathbf{q}}(t) \in \mathbb{Q} \triangleq [\mathcal{Q}_1, \mathcal{Q}_L]^{\mathcal{N}}$, where $\mathbb{I}^+$ is the set of positive integers.
Finally, the state space can be defined as $\mathbb{S} \triangleq \mathbb{P}\times \mathbb{C} \times \mathbb{B} \times \mathbb{H} \times \mathbb{Q}$.

\subsubsection{\textbf{Action}}
As mentioned earlier, for supporting adaptive video streaming, the MBS decides three actions as follows: 1) which chunk, 2) what quality, and 3) how many chunks would be provided from the MBS to each mBS, depending on the states explained above.
The action of the proposed MDP at slot $t$ is defined as $a_t \triangleq \{\mathbf{l}(t) \}$, where $\mathbf{l}(t) = [\mathbf{l}_j^i : x_j \in \mathcal{K}, u_i \in \mathcal{U}_j]^T$. 
Here, $\mathbf{l}(t) \in \mathbb{L} \triangleq \{0,1,\cdots,\bar{c}\}^{\mathcal{K} \cdot \mathcal{N} \cdot L}$. 
Therefore, the action space can be defined as $\mathbb{A} \triangleq \mathbb{L}$.

\subsubsection{\textbf{Random event}}
The time-varying wireless channel conditions between mBSs and vehicles are the random event in this MDP. 
Even though we focus on pushing video chunks from the MBS to mBSs, queue and buffer states of mBSs and vehicles separately are dependent on the content delivery from mBSs to vehicles; therefore, the random distribution of wireless channels between mBSs and vehicles should be considered in the proposed MDP. 
The channel capacity of the link between mBS $x_j$ and vehicle $u_i$ are given randomly for all $x_j \in \mathcal{X}$ and $u_i \in \mathcal{U}_j$, denoted by $r_{i,j}(t)$. 
Accordingly, the size of all the transmitted chunks from mBS $x_j$ to vehicle $u_i$ satisfying $p_i(t) \in \{j-1, j\}$ at slot $t$ should be upper bounded on $r_{i,j}(t)$, as given by
\begin{equation}
\sum_{q_l\in \mathcal{Q}} \tilde{C}_l w_{j}^{(i,q_l)}(t) \leq r_{i,j}(t).
\end{equation}

Since it is difficult for the MBS to collect all the channel information between mBSs and vehicles in a short time and to track the channel of vehicles with high mobility, the MBS learns the streaming policy without the knowledge of channel gains. 
Denote a probability mass function of the actual number of chunks of quality $q$ that can be delivered from mBS $x_j$ to vehicle $u_i$ at slot $t$ depending on randomly generated $r_{i,j}(t)$ by $f_j^{(i,q)}(t,w)$. 
Since vehicles are moving so that distances between mBSs and vehicles are time-varying, $f_j^{(i,q)}(t,w)$ could be non-stationary; therefore, the dependency of $t$ is given.
Note that this does not mean that the data rate of each user has a Markov property; however, the data rate depends on the random channel conditions of vehicles, distances between mBSs and vehicles related to pathloss effects, and any other network randomness.
In this paper, a uniform distribution is assumed for $r_{i,j}(t)$ for simplicity, i.e., $r_{i,j} \sim \text{unif}(0,r_{\text{max}})$ where $r_{\text{max}}$ is the upper bound on $r_{i,j}(t)$, if $u_i$ is in the coverage of $x_j$, i.e., $p_i(t) = j$. 

\textit{Remark}: Note that the data rate $r_{i,j}$ can follow any distribution and the proposed adaptive streaming scheme can be applied to any distribution of $r_{i,j}(t)$, not limited to the uniform distribution. 
In fact, distribution of $r_{i,j}(t)$ could incorporate any randomness of the network (i.e., random channel distribution, pathloss, Doppler effects, power control, bandwidth allocation); therefore, the proposed DRL agent enables to train the streaming policy in any random network environment and/or with any random resource allocation rule.
Since the learning agent only needs to observe the updated queue and buffer states of mBSs and vehicles respectively without the knowledge of each element of network randomness, we do not have to assume a specific distribution of $r_{i,j}(t)$ and to investigate the impacts of Doppler effects or resource allocation on our policy.
If the content delivery scheme from mBSs to vehicles and resource allocation for links between mBSs and vehicles are considered, their dependency on the distribution of $r_{i,j}(t)$ should be investigated; however, this is out of scope for this manuscript and we will study in the future work. 

\subsubsection{\textbf{State transition}}
The state transition probability and its mechanism are the main factors for determining how to learn the policy and how the policy is converged.
Denote the current and next states by $s = \{ \mathbf{p}, \mathbf{c}, \mathbf{b}, \mathbf{h}, \bar{\mathbb{q}} \}$ and $s' = \{ \mathbf{p}', \mathbf{c}', \mathbf{b}', \mathbf{h}', \bar{\mathbb{q}}' \}$, respectively.
First, the mBS association state follows the FSMC model and does not depend on the action or the random event.
Therefore, the state transition probability of $\mathbf{p}$ is given by
\begin{equation}
P_{\mathbf{p},\mathbf{p'}} = \prod_{u_i \in \mathcal{U}} \alpha_i,~\text{where}~\alpha_i = 
\begin{cases}
\rho & \text{if}~p'_i = p_i+1 \\
1-\rho & \text{elseif}~p'_i = p_i \\
0 & \text{otherwise}
\end{cases}.
\end{equation}

Second, queue and buffer states are influenced by the action $\boldsymbol{l}$ and the random event $\mathbf{w}$.
Given the action $\boldsymbol{l}$, the state transits from $\mathbf{c}$ to $\mathbf{c}'$ with the probability of 
\begin{align}
&~~P_{\mathbf{c},\mathbf{c}'}(\boldsymbol{l}) = \prod_{\substack{x_j \in \mathcal{X} \\ u_i \in \mathcal{U}, q\in\mathcal{Q}}} \beta_{j}^{(i,q)},~\text{where} \nonumber \\ 
&\beta_{j}^{(i,q)} = 
\begin{cases}
f_{j}^{(i,q)}(l_{j}^{(i,q)} - c_{j}^{(i,q)'} + c_{j}^{(i,q)}) &  \text{if } c_{j}^{(i,q)'} \in \mathcal{C}_j^{(i,q)} \\
0 & \text{otherwise}
\end{cases},
\label{eq:p_st_c}
\end{align}
where $\mathcal{C}_j^{(i,q)} = [l_j^{(i,q)}, c_j^{(i,q)}+l_j^{(i,q)}]$.
On the other hand, the buffer state $b_i$ of the vehicle $u_i$ is influenced by the summation of $w_{p_i(t)}^{(i,q)}$ values for all $q\in \mathcal{Q}$. 
According to the `high-quality first' rule for delivering chunks from mBSs to vehicles explained in Section \ref{subsec_IV_II}, the state transition probability of $\mathbf{b}$ is obtained as \eqref{eq:p_st_b} at the top of this page.
\begin{figure*}
	\begin{align}
	&~~P_{\mathbf{b},\mathbf{b}'}(\boldsymbol{l} | \mathbf{c}) = \prod_{u_i \in \mathcal{U}} \varsigma_i,~\text{where}~\nonumber \\
	&\varsigma_i = 
	\begin{cases}
	f_j^{(i,L)}(b'_i - \tilde{b}_i) & \text{if } \tilde{b}_i \leq b'_i \leq \tilde{b}_i + c_j^{(i,L)} \\
	f_j^{(i,L)}\big(c_j^{(i,L)} \big) \cdot f_j^{(i,L-1)} \big( b'_i - \tilde{b}_i - c_j^{(i,L)} \big) & \text{elseif } \tilde{b}_i + c_j^{(i,L)} < b'_i \leq \tilde{b}_i + c_j^{(i,L)} + c_j^{(i,L-1)} \\
	~~~~~~~~~~~~~~~~\vdots & ~~~~~~~~~~~~~~~~\vdots \\
	\prod_{q=l+1}^L f_j^{(i,q)}\big( c_j^{(i,q)} \big) \cdot f_j^{(i,l)}\big( b'_i - \tilde{b}_i - \sum_{q=l+1}^L c_j^{(i,q)} \big) & \text{elseif } \tilde{b}_i + \sum_{q=l+1}^L c_j^{(i,q)} < b'_i \leq \tilde{b}_i + \sum_{q=l}^L c_j^{(i,q)} \\
	~~~~~~~~~~~~~~~~\vdots & ~~~~~~~~~~~~~~~~\vdots \\
	\prod_{q=2}^L f_j^{(i,q)}\big( c_j^{(i,q)} \big) \cdot f_j^{(i,1)}\big( b'_i - \tilde{b}_i - \sum_{q=2}^L c_j^{(i,1)} \big) & \text{elseif } \tilde{b}_i + \sum_{q=2}^L c_j^{(i,q)} < b'_i \leq \tilde{b}_i + \sum_{q=1}^L c_j^{(i,q)} \\
	0 & \text{otherwise}
	\end{cases}
	\label{eq:p_st_b}
	\end{align}
	\hrulefill
\end{figure*}
In \eqref{eq:p_st_b}, $\tilde{b}_i = \max\{b_i - \mathcal{F}, 0 \}$.
Note that $P_{\mathbf{b},\mathbf{b}'}$ depends on the current transmission queue state $\mathbf{c}$, because the number of delivered chunks (i.e., $w_j^{(i,q)}$) to the receiver buffer cannot exceed the number of chunks accumulated in mBS queues (i.e., $c_j^{(i,q)}$).
In addition, the states of $h_i$ and $\bar{q}_i$ transit to $h'_i$ and $\bar{q}'_i$ depending on the accumulated sums of $w_{p_i(t)}^{(i,q)}$ values for all $q\in \mathcal{Q}$; therefore, their state transition probabilities can be computed in a similar way of \eqref{eq:p_st_b}. 

\begin{algorithm}[t]
	Initialize the \textit{critic} and \textit{actor} networks with weights $\theta^{Q}$ and $\theta^{\mu}$~\\
	Initialize the target networks as: $\theta^{\hat{Q}} \leftarrow \theta^{Q}, \theta^{\hat{\mu}} \leftarrow \theta^{\mu}$ ~\\
	\For{episode = 1, $\mathcal{E}$}{
		\textbf{Phase 1. Initialize the replay buffer $\Phi$}~\\ 
		\For{mini batch = 1 to c}{
			$\triangleright$ Randomly generate $\varphi$ states $s \in \mathbb{S}$ ~\\
			$\triangleright$ Get corresponding a set of actions $a = \mu(s|\theta^{\hat{\mu}}) \in \mathbb{A}$ for each $s$~\\
			$\triangleright$ Input the state-action pairs to predefined \textbf{vehicular network environments} and get a set of reward $r$ for each pair based on~(\ref{eq:reward_total}), and observe the next set of states $s \in \mathbb{S}^{'}$}
		$\triangleright$ Store the transition pairs $\xi = (s, a, r, s^{'})$ as a minibatch $\phi$, which composes the $\Phi$.
	} 
	\textbf{Phase 2. Update target networks periodically~\\}
	\For{time step = 1, $\mathcal{T}$}{
		\textbf{If} \textit{time step} \textbf{is update period, do followings:}~\\
		$\triangleright$ Sample a random minibatch without replacement from $\Phi$~\\
		$\triangleright$ Set $y_{i} = r_{i} + \gamma \hat{Q}(s_{i}^{'}, \mu(s_{i}^{'}|\theta^{\hat{\mu}})|\theta^{\hat{Q}})$ ~\\
		$\triangleright$ Update the $\theta^{Q}$ by applying stochastic gradient descent to the loss function of \textit{critic} network, which can be obtained as $\frac{1}{\varphi}\sum_{i}{(y_{i} - Q(s_{i}, a_{i}|\theta^{Q}))}^{2}$ ~\\
		$\triangleright$ Update the $\theta^{\mu}$ by applying stochastic gradient ascent with respect to the gradient of \textit{actor} network: ~\\
		$\nabla_{\theta^{\mathcal{A}}}J(\theta^{\mu}) \approx \frac{1}{\varphi}\sum_{i}{\nabla_{a}Q(s, a|\theta^{Q})\nabla_{\theta^{\mu}}\mu(s|\theta^{\mu})|_{s=s_{i},a=\mu(s_{i}|\theta^{\mu})}}$~\\
		$\triangleright$ \textit{Soft} update $\theta^{\hat{Q}}$ and $\theta^{\hat{\mu}}$ as~(\ref{target_update})}
	\caption{DDPG-based adaptive video streaming algorithm for mmWave vehicular networks}
	\label{algorithm1}
\end{algorithm}

\subsubsection{\textbf{Reward}}
At every time slot $t$, the learning agent $\mathcal{M}$ observes the current state $s_t$ and determines the action $a_t$, then the state is changed to $s_{t+1}$. 
In this step, the agent acquires the system reward $r_{t+1}$, which includes a composite reward of (i) smooth quality enhancement, (ii) avoiding packet drop, and (iii) seamless playback without stall events.
These sub-rewards are denoted as $R^{q}(s_t, a_t)$, $R^{p}(s_t, a_t)$, and $R^{f}(s_t, a_t)$, respectively, and the reward structure is designed as $r_{t+1} = R(s_t, a_t) = R^q(s_t, a_t) \otimes R^p(s_t, a_t) + R^f(s_t, a_t)$. 
Each sub-reward is the summation of rewards corresponding to all vehicles, as given by
\begin{align}
R(s_t, a_t) 
&= \overline{r^q(t)} \cdot \overline{r^p(t)} + \overline{r^f(t)},
\label{eq:reward_total}
\end{align}
where $\overline{r^q(t)} = \frac{1}{\mathcal{N}} \sum_{i=0}^{\mathcal{N}-1} r_i^q (t)$, $\overline{r^p(t)} = \frac{1}{\mathcal{K}} \sum_{j=0}^{\mathcal{K}-1} r_j^p (t)$, and $\overline{r^f(t)} = \frac{1}{\mathcal{N}} \sum_{i=0}^{\mathcal{N}-1} r_{i}^{f}(t)$.

The reward of smooth quality enhancement, i.e., $R^q(s_t, a_t)$, basically pursues the high-quality chunks while avoiding the frequent and steep quality fluctuations. 
Since sudden changes in video quality could be detrimental to users' QoS, even when the streaming system is sufficient to provide the highest-quality, it is better to monotonically increase the quality.
Similarly, when the channel condition is not good so that the mBS tries to deliver the low-quality chunks, the agent slowly decreases the video quality. 
Then, we have four different user's satisfaction cases as follows:\\\\
\begin{tabular}{ll}
	\textit{Case 1.}& Quality is high and quality fluctuation is high.\\
	\textit{Case 2.}& Quality is high and quality fluctuation is low.\\
	\textit{Case 3.}& Quality is low and quality fluctuation is high.\\
	\textit{Case 4.}& Quality is low and quality fluctuation is low.\\ 
\end{tabular}\\\\
Here, $R^q(s_t, a_t)$ is designed to maximize \textit{Case 2}, which means to promote increasing the average video quality while limiting quality fluctuations. 
Therefore, the quality reward for each vehicle $u_i$ can be modeled as follows:
\begin{equation}
r_i^q(t) = \frac{\frac{1}{\sum_{q\in\mathcal{Q}} w_{p_i(t)}^{(i,q)}(t)} \sum_{q\in\mathcal{Q}} q\cdot w_{p_i(t)}^{(i,q)}(t)}{\epsilon + \big| \bar{q}_i - \frac{1}{\sum_{q\in\mathcal{Q}} w_{p_i(t)}^{(i,q)}(t)} \sum_{q\in\mathcal{Q}} q\cdot w_{p_i(t)}^{(i,q)}(t) \big|},
\label{eq:r_i^q}
\end{equation}
where $\frac{1}{\sum_{q\in\mathcal{Q}} w_{p_i(t)}^{(i,q)}(t)} \sum_{q\in\mathcal{Q}} q\cdot w_{p_i(t)}^{(i,q)}(t)$ indicates the average quality of the chunks arrived at the vehicle in the current slot, and $\epsilon$ is a system parameter that weighs the importance of the control of quality fluctuations. 
Note that the numerator of \eqref{eq:r_i^q} represents the current video quality and the denominator of \eqref{eq:r_i^q} measures the current quality fluctuation.

By pursuing rewards $R^p(s_t, a_t)$ and $R^f(s_t, a_t)$, the adaptive streaming policy is willing to avoid packet drop and playback stall events, respectively.
When the pushed chunks from the MBS to mBS $x_j$ make the transmitter queue of $x_j$ overflow, some of the pushed chunks should be discarded because the queue length is upper bounded on $\bar{c}$.
Accordingly, $r_j^p$ can be designed as the ratio of quantities of the discarded chunks to the pushed chunks, as given by
\begin{equation}
    r_j^p(t) = \sum_{u_i\in \mathcal{U}} \sum_{q\in \mathcal{Q}} r_{j,(i,q)}^p(t),
\end{equation}
where
\begin{equation}
r_j^p(t) = ( 1- \eta) - \frac{ \max\{\tilde{c}_j^{(i,q)}(t) + l_j^{(i,q)}(t) - \bar{c}, 0\} }{\zeta + | l_j^{(i,q)}(t) |}.
\label{eq:r_i^p}
\end{equation}
In \eqref{eq:r_i^p}, $\zeta$ and $\eta$ are parameters that scale the reward for packet drops.
The numerator of \eqref{eq:r_i^p} indicates the number of dropped chunks due to the finite queue size and the denominator of \eqref{eq:r_i^p} is the number of chunks delivered from the MBS to the queue of mBS $x_j$.
Therefore, $r_j^p(t)$ measures how much portion of $l_j^{(i,q)}(t)$ is dropped at the queue of mBS $x_j$. 

Lastly, $r_i^f$ measures the penalty given when the buffer length of a vehicle $u_i$ is not sufficient to play the fixed number of chunks during one time slot (i.e., $\mathcal{F}$).
Therefore, $r_i^{f}(t)$ can be designed by
\begin{equation}
r_i^f(t) = \nu \cdot \min\{ (b_i(t) - \mathcal{F}), 0 \}, \label{eq:r_i^f}
\end{equation}
where $\nu > 0$ represents the scaling factor for $r_i^f(t)$.
In \eqref{eq:r_i^f}, $r_i^f(t) \leq 0$, and we can see that the negative reward is given if the playback stall event happens.

The structure of the reward function in \eqref{eq:reward_total} is obtained by experiments in which the function in \eqref{eq:reward_total} makes the learning process converge well and balances the tradeoff among conflicting performance metrics, i.e., the average quality, the average quality variations, packet drop rates, and playback stall rates. 
Nevertheless, we can intuitively explain that packet drops scale the quality reward. 
Specifically, when the packet drop event happens, it means that the backhaul is largely consumed for providing the high-quality chunks with large file sizes. 
In this case, even vehicle users enjoy the high quality streaming with small quality variations, its large quality reward is obtained at the expense of excessive backhaul usage; therefore, the quality reward is scaled down by multiplying the penalty of packet drops.

\subsubsection{\textbf{Algorithm}}
Based on this MDP structure, we finally formulate the optimization problem to maximize the reward \eqref{eq:reward_total} during $T$ time slots as follows:
\begin{align}
&\underset{ \mathbf{a} =[a_1, \cdots, a_T] }{\arg\max}~\sum_{t=1}^T R(s_t,a_t) \label{eq:opt_prob} \\
&~~\text{s.t.}~c_{j}^{(i,q)}(t) \leq \bar{c},~\forall x_j ,~\forall u_i,~\forall q,~\forall t \label{eq:opt_const1} \\
&~~~~~~~b_i(t) \leq \bar{b},~\forall u_i \label{eq:opt_const2} \\
&~~~~~~~\bar{q}_i(t) \in \mathcal{Q},~\forall u_i. \label{eq:opt_const3}
\end{align}
Note that in a realistic system with massive number of vehicles on the highway, the system state space is very extensive and it is very difficult for the MBS to handle and compute all the network states. 
In addition, the average quality state $\bar{\mathbf{q}}$ is continuous. 
Accordingly, the problem of \eqref{eq:opt_prob}--\eqref{eq:opt_const3} cannot be solved by a classical method without the help of neural networks, e.g., dynamic programming; thus, this justifies the use of the DDPG-based DRL approach as proposed in this paper. 
As mentioned earlier, the MBS is the learning agent and performs Algorithm \ref{algorithm1} in order to learn the action of the proposed MDP for adaptive video streaming that can improve its quality while limiting packet drops, playback stall events, and excessively large quality fluctuations.
The learning procedures of the adaptive streaming policy are as follows:

\begin{itemize}
	\item First, the parameters of the \textit{actor} and \textit{critic} networks, which activate and evaluate the action of $\mathcal{M}$, are initialized (line 1).
	\item Then, the target networks for both the \textit{actor} and \textit{critic} network, $\hat{\mu}$ and $\hat{Q}$, are initialized (line 2).
	\item By iterating each episode, $\mathcal{M}$ repeats the following learning procedures: 
	\subitem (i) for every episode, $\varphi$ transition pairs including the set of the randomly generated $\varphi$ states (i.e., $\mathcal{S} \in \mathbb{S}^{\varphi}$), the possible action sets given $\mathcal{S}$ (i.e., $\mathcal{A} \in \mathbb{A}^{\varphi}$), the reward set for $\mathcal{S}$ and $\mathcal{A}$, and the observed next state set (i.e., $\mathcal{S}^{'} \in \mathbb{S}^{\varphi}$) are collected as \textit{minibatch} $\phi$, and they are stored in \textit{replay buffer} $\Phi$ (lines 4--9). 
	
	\subitem (ii) After $\Phi$ is fully constructed, a minibatch in $\Phi$ is randomly sampled without replacement. Then, the $i$-th transition pair of the minibatch is utilized for calculating the mean squared Bellman error (MSBE) between the target value $y_{i}$ and $Q(s_{i}, a_{i}|\theta^{Q})$, i.e., $\frac{1}{\varphi}\sum_{i}{(y_{i} - Q(s_{i}, a_{i}|\theta^{Q}))}^{2}$, in order to update the \textit{critic} network with the gradients obtained from the difference (line 14--15). In addition, gradient of $\theta^{\mu}$ is utilized to update the parameters of the \textit{actor} network (line 16--17). Overall, the updated parameters of the \textit{critic} and \textit{actor} networks are utilized to update the target parameters $\theta^{\hat{\mathcal{Q}}}$ and $\theta^{\hat{\mu}}$ with \textit{soft} update weight $\tau$ as (\ref{target_update}) to ensure efficient and stable learning (line 11--18) until convergence. 
\end{itemize}

	\section{Performance Evaluation and Discussions}\label{sec:6}
This section presents simulation results to verify the performance of our DDPG-based adaptive streaming system in mmWave-based vehicular networks. 
We evaluate the proposed streaming system with the aforementioned rewards given that the state of the vehicular network is observable by the agent.
We leveraged Pytorch in our simulations to implement our proposed DDPG-based adaptive streaming system. 
The simulation settings are first presented and then we discuss the simulation results.

\subsection{Simulation Settings}
\label{subsec_V_I}
In this subsection, we elaborate on the implementation details of the proposed DDPG learning-based adaptive streaming system for an mmWave vehicular network. We first introduce the hardware configuration for our simulation, and then we present the overall design and implementation details of the software. In terms of hardware, we employed a NVIDIA DGX station equipped with 4 $\times$ Tesla V100 GPUs (a total of 128 GB memory available) and an Intel Xeon E5--2698 v4 2.2 GHz CPU with 20 cores (256 GB system memory available in total). 
Next, in terms of software, we utilized Python version 3.6 on Ubuntu 16.04 LTS to build the DDPG-based adaptive video streaming scheme. In addition, we used a fan-in initializer and Xavier initializer~\cite{Xavier} to avoid the occurrence of vanishing gradient descent during the learning phase. The neural network was constructed with a fully connected deep neural network (DNN), and four hidden layers are used. 
Each of the hidden layers has 500, 400, 300 or 200 nodes. 
The reinforcement learning model was trained through a total of 500,000 iterations. 

According to \cite{ICTC2015Kim}, $\mathcal{Q}=\{q_1=1, q_2=2.5, q_3=5, q_4= 10, q_5 = 25\}$ Mbps are used for the bitrates of videos depending on quality levels.
In addition, $L=5$ and $\tau=1$ are assumed; therefore, $q_l = \tilde{C}_l$ for all $l\in\{1,2,3,4,5 \}$.  
The parameters for learning procedures are summarized in Table.~\ref{tab:tab2}

We design two main comparison techniques (i.e., random action and proactive caching proposed in~\cite{TWC2016Qiao}) as the baseline. 
The random action technique is that MBS conducts cache allocation to mBS with the limited cache bound (e.g. $0.5\bar{c}$, and $1.0\bar{c}$). 
Note that the comparison technique with cache bound $0.5\bar{c}$ and with $1.0\bar{c}$ are associated with \textit{Comp1}, respectively. Second, we compare our proposed model to a proactive caching technique, which is state of the art and is denoted as \textit{Comp2}.

\begin{table}[t]
	\caption{Experiment Parameters}
	\footnotesize
	\label{tab:tab2}
	\begin{center}
		\centering
		\begin{tabular}{c||l||l}
			\toprule[1.0pt]\centering
			Parameter & Value & Description\\
			\midrule
			$\epsilon$ & 1.0 & The parameter of quality fluctuation reward \\
			$\zeta$ & 0.05 & The parameter of cache overflow reward \\
			$\eta$ & 1.0 & The parameter of cache overflow reward \\
			$\nu$ & 1.0 & The parameter of playback stall reward \\
			$b^c$ & 2.0 &  The consumption of buffer length \\
			\midrule
			$\gamma_{a}$ & $3\times10^{-4}$ & The learning rate of actor network\\			
			$\gamma_{c}$ & $3\times10^{-4}$ & The learning rate of critic network\\
			$\tau$ & $10^{-2}$ & Target smoothing coefficient\\
			$\mid \Phi \mid$ & $1000$ & Batch size \\
			\midrule
			$\mathcal{K}$ & 20 & The number of mBSs \\
			$\mathcal{N}$ & 200 & The number of vehicles \\
			$\bar{c}$ & 2400 & The length of transmitter queues of mBSs \\
			$\bar{b}$ & 240 & The length of receiver buffers of vehicles \\
			$\mathcal{F}$ & 40 & The playback rate \\
			\bottomrule[1.0pt]
		\end{tabular}
	\end{center}
\end{table}

\begin{figure*}[t!]
\centering \small
\setlength{\tabcolsep}{2pt}
\renewcommand{\arraystretch}{0.2}
\begin{tabular}{p{0.25\linewidth}p{0.25\linewidth}p{0.25\linewidth}p{0.25\linewidth}}
\includegraphics[page=1, width=1\linewidth]{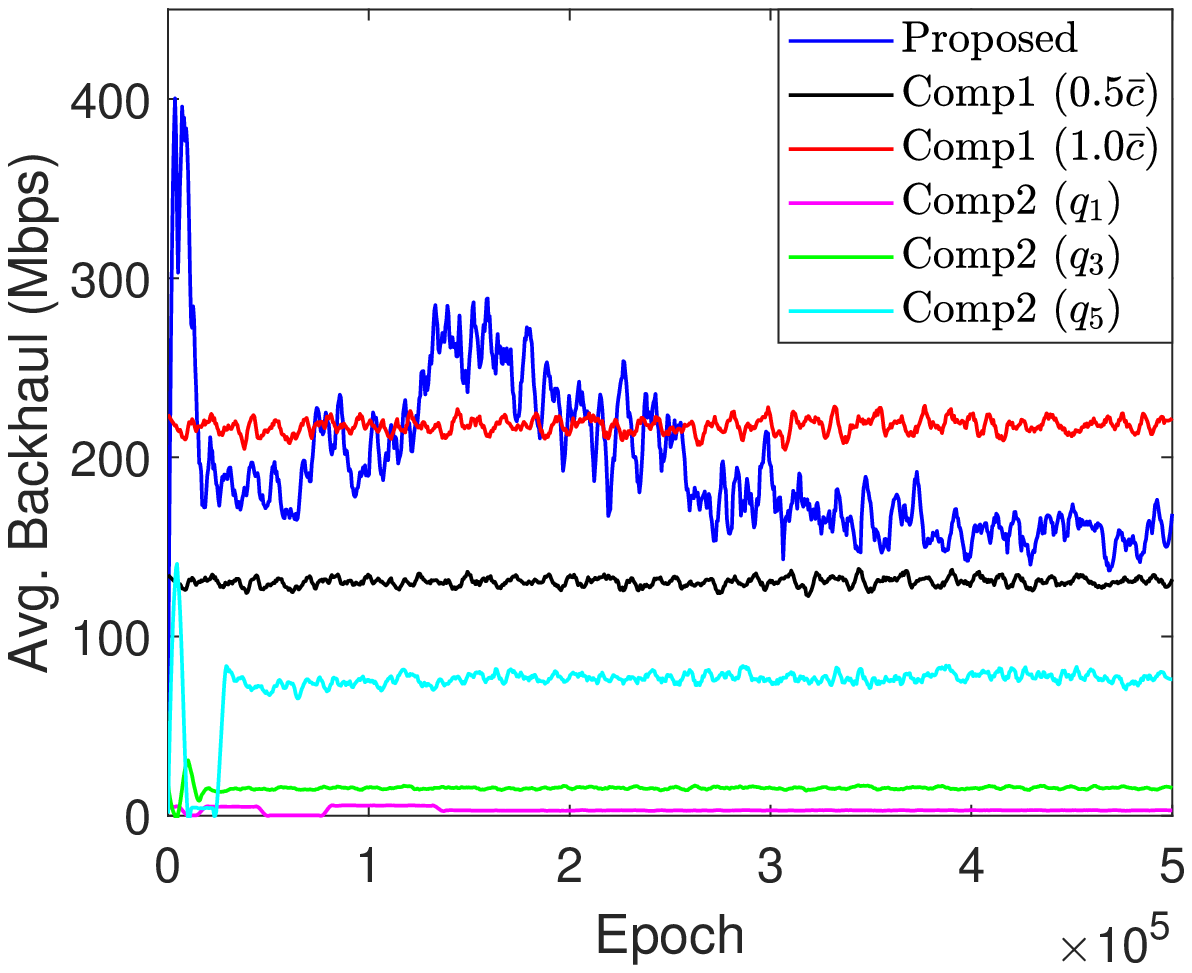} & 
\includegraphics[page=1, width=1\linewidth]{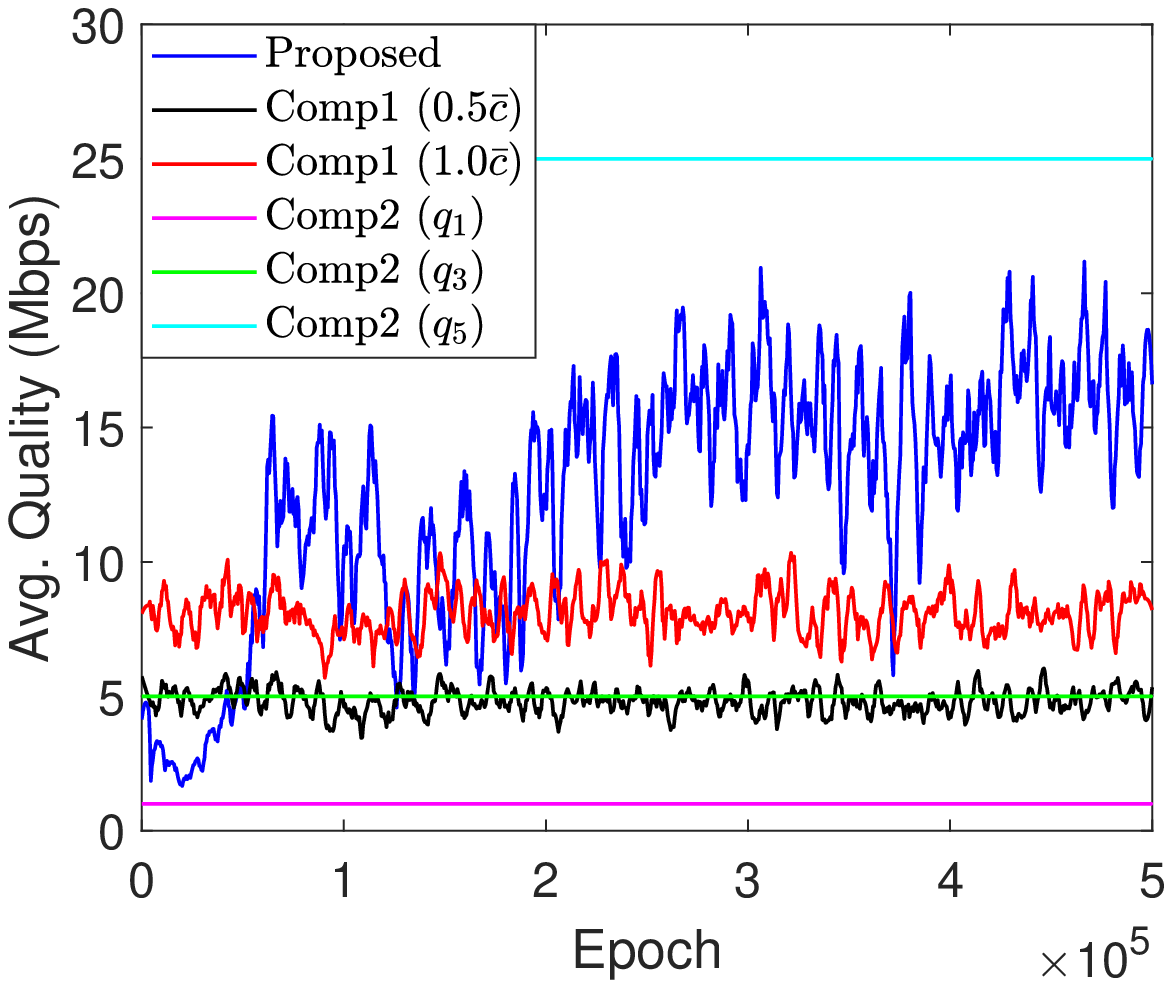} &
\includegraphics[page=1, width=1\linewidth]{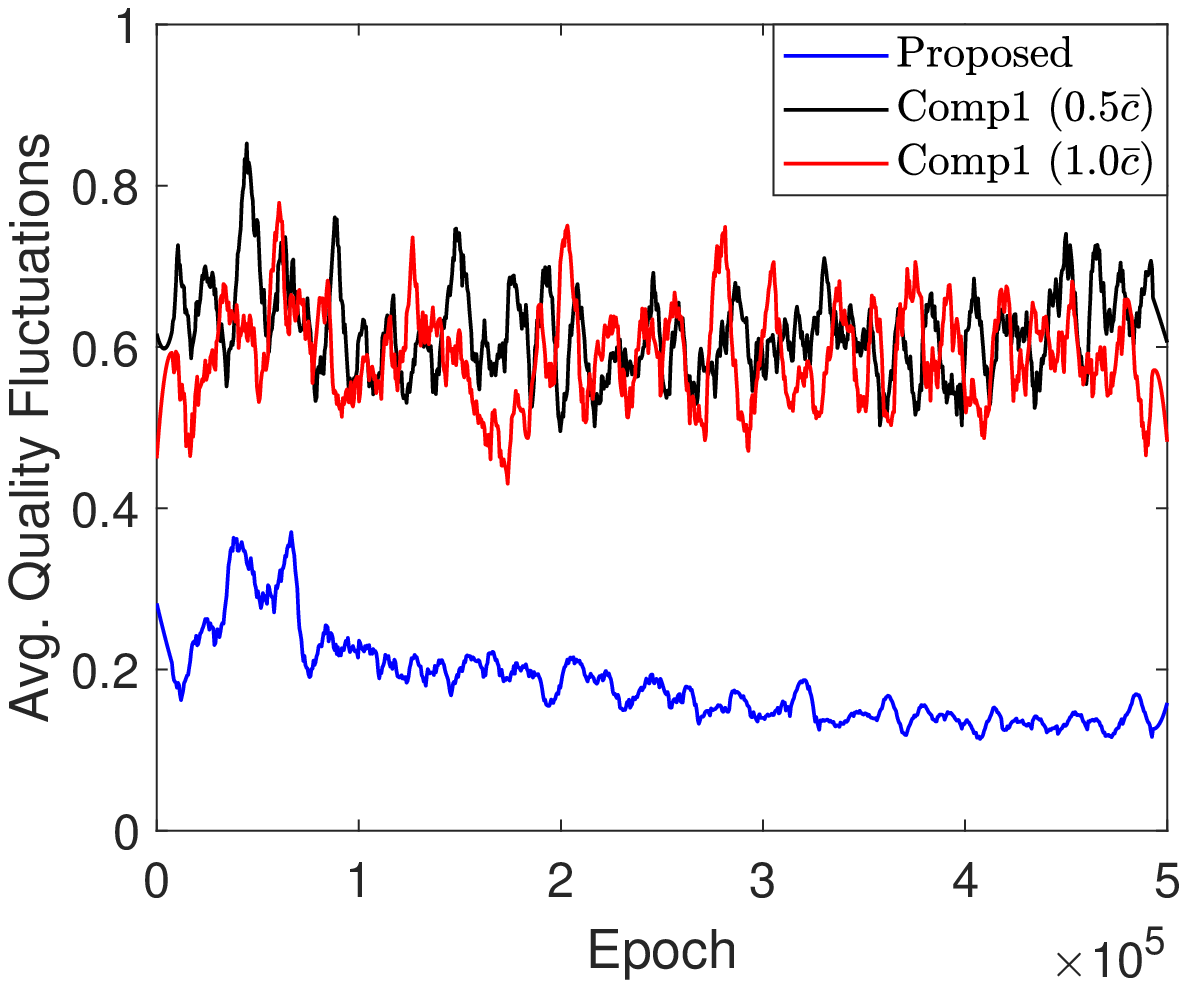} & 
\includegraphics[page=1, width=1\linewidth]{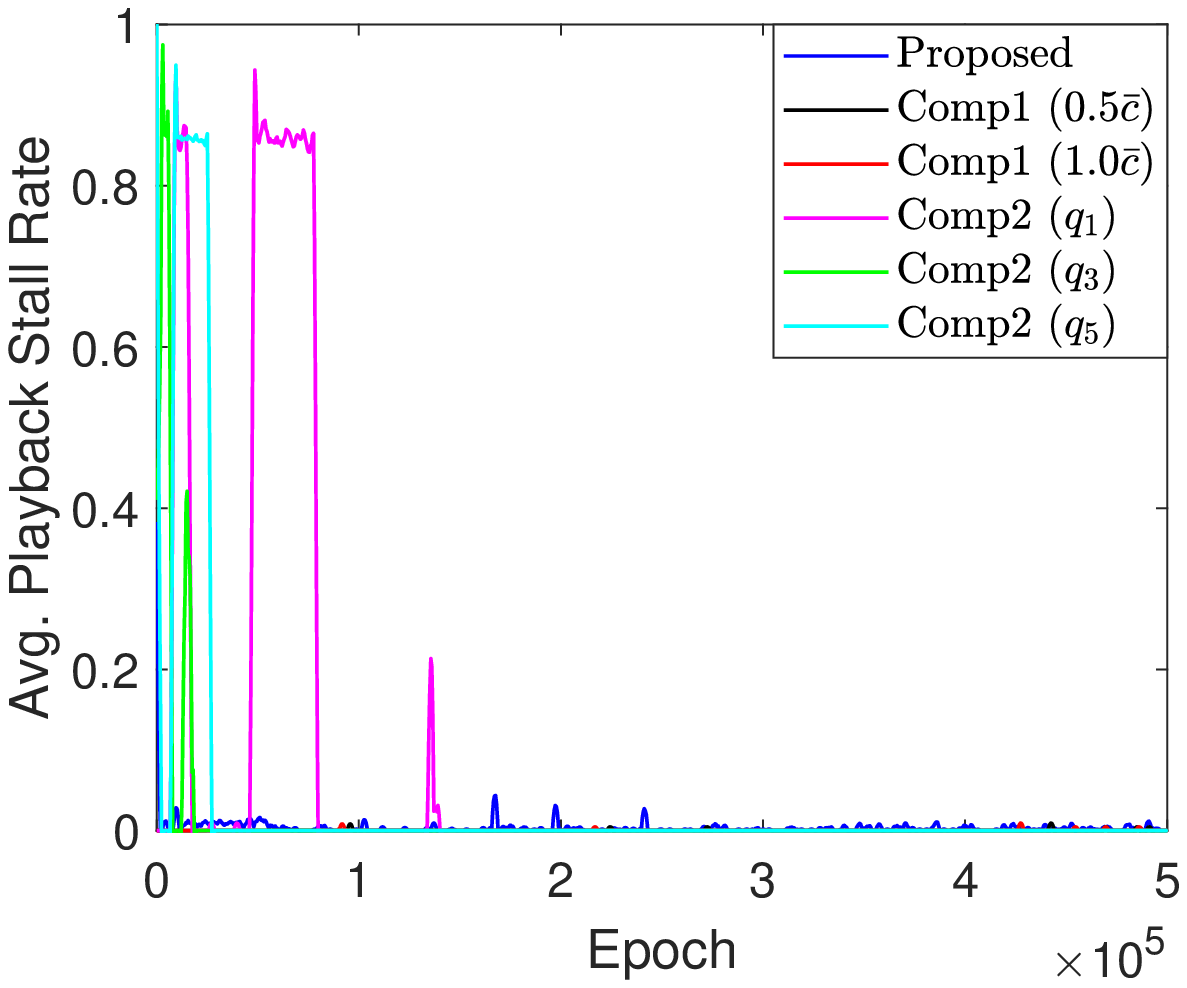}
\tabularnewline
\tabularnewline
\small\centering(a) Backhaul usage.&
\small\centering(b) Quality.&
\small\centering(c) Quality variations. &
\small\centering(d) Playback stall.\\
\tabularnewline
\includegraphics[page=1, width=1\linewidth]{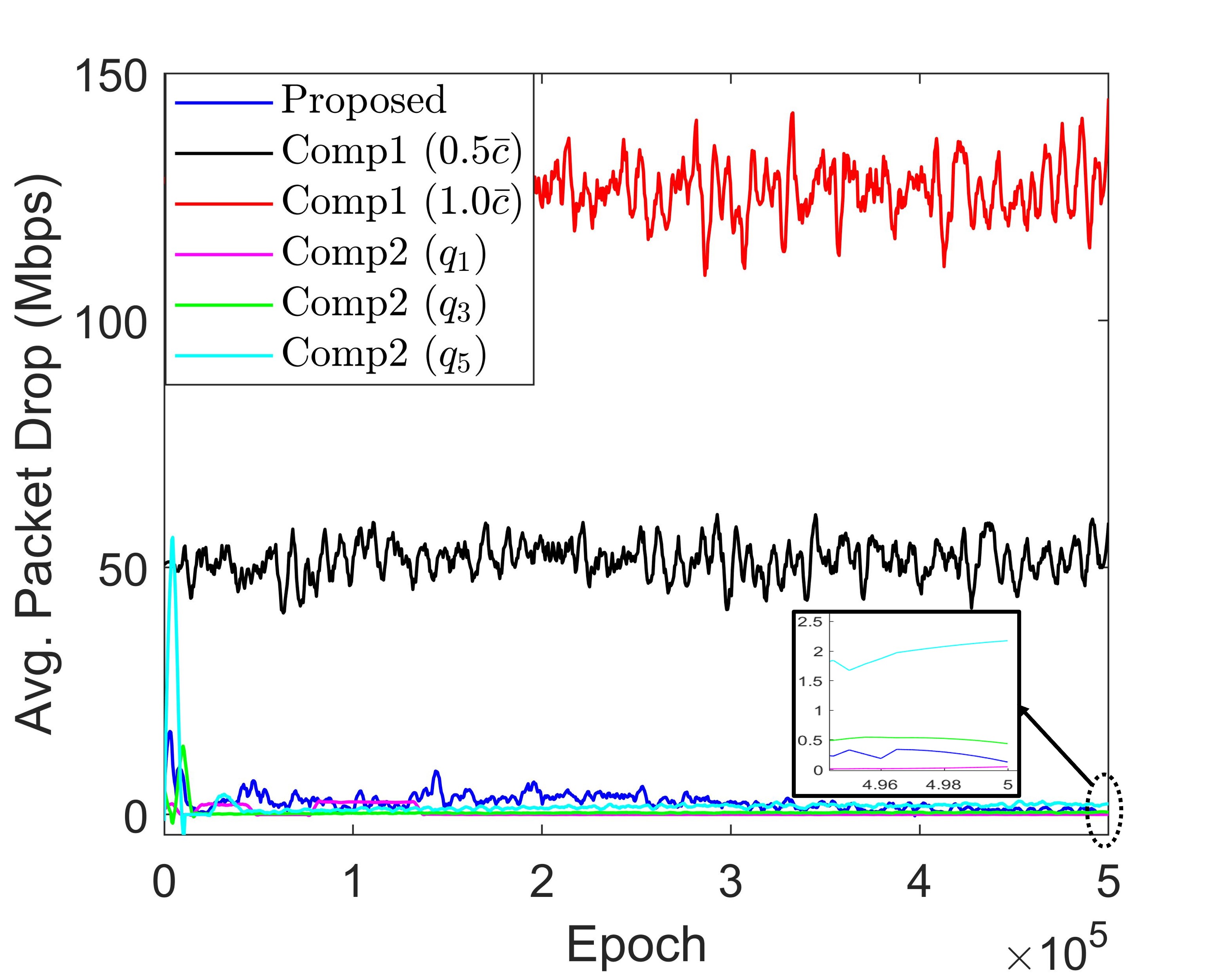}& 
\includegraphics[page=1, width=1\linewidth]{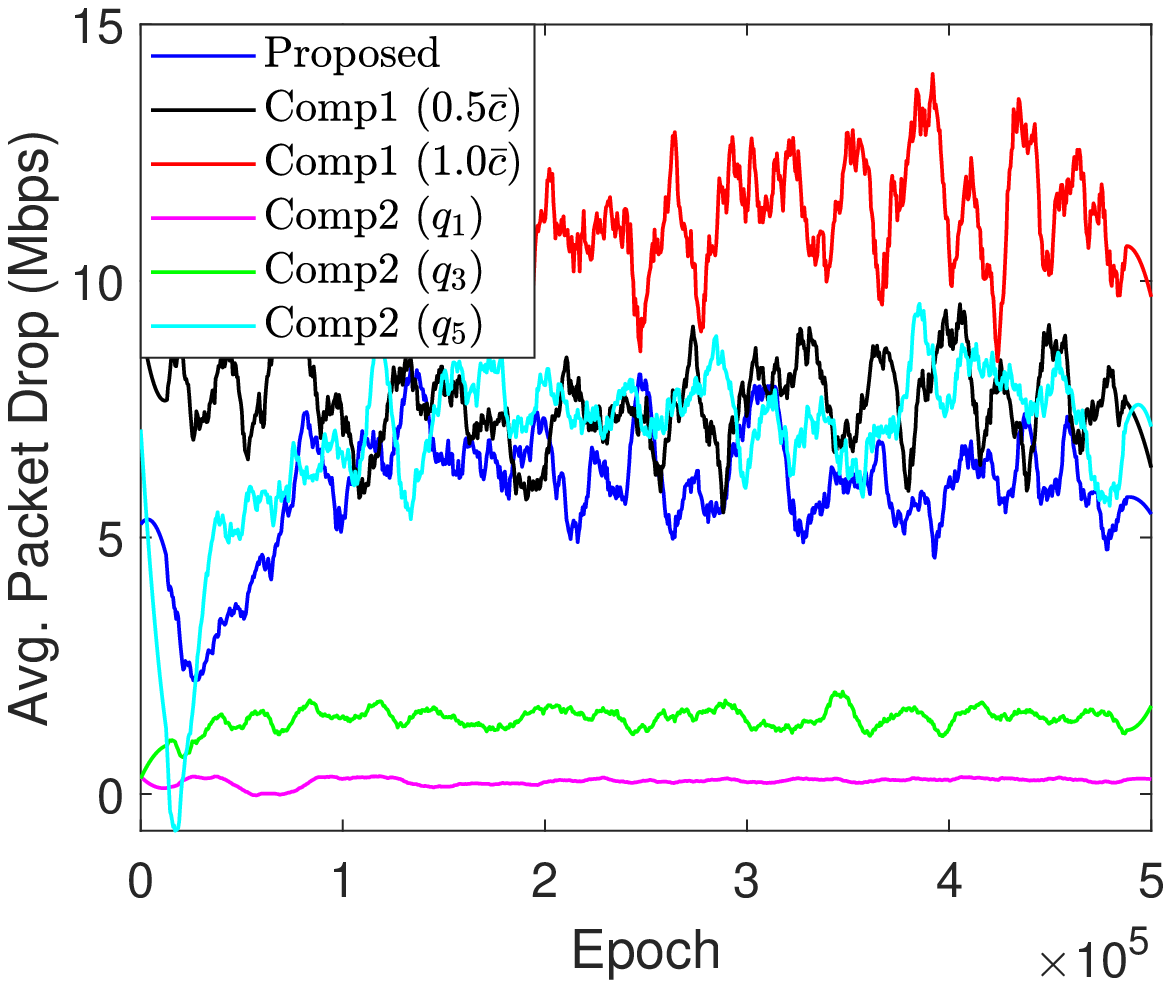} & 
\includegraphics[page=1, width=1\linewidth]{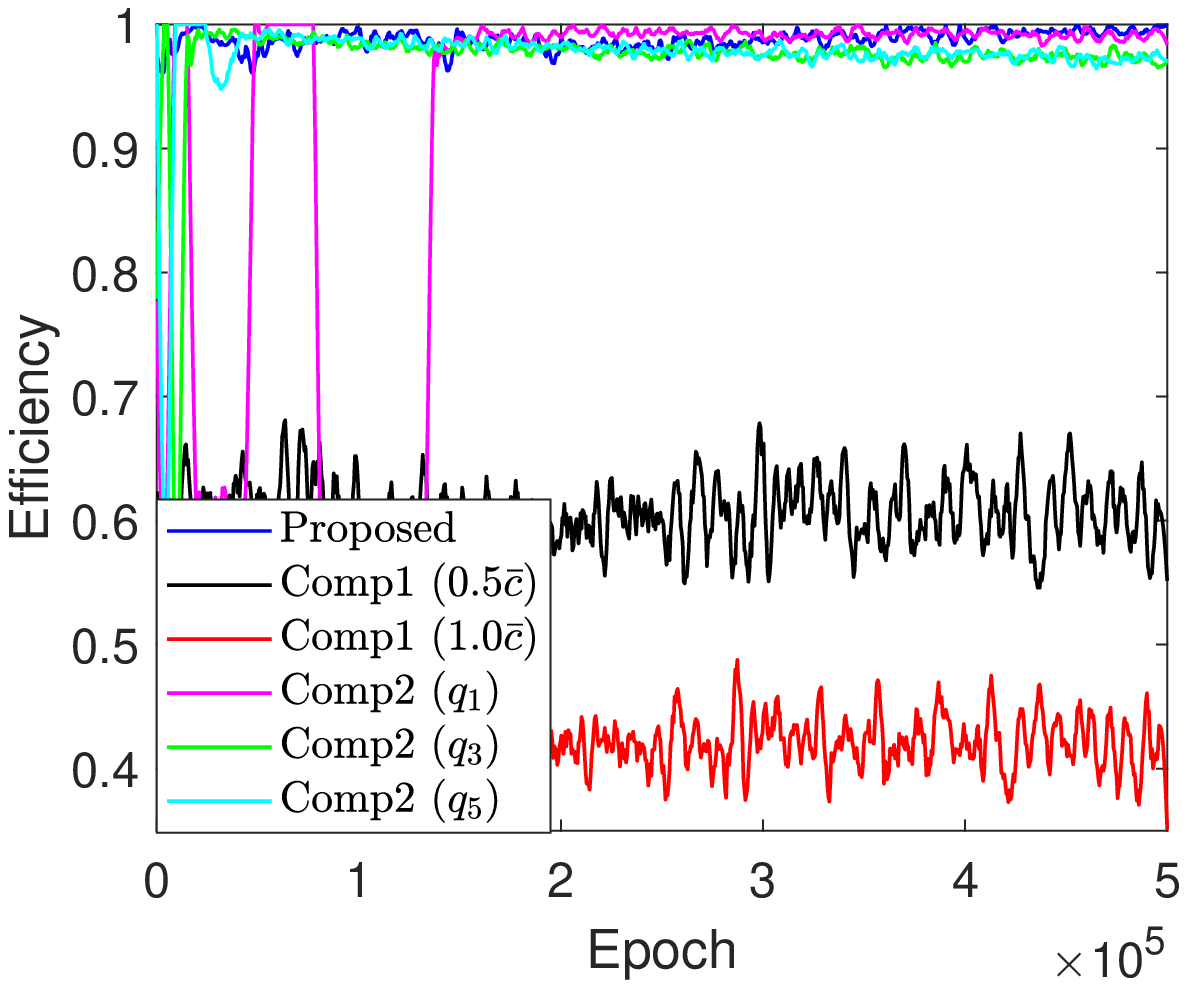}&
\includegraphics[page=1, width=1\linewidth]{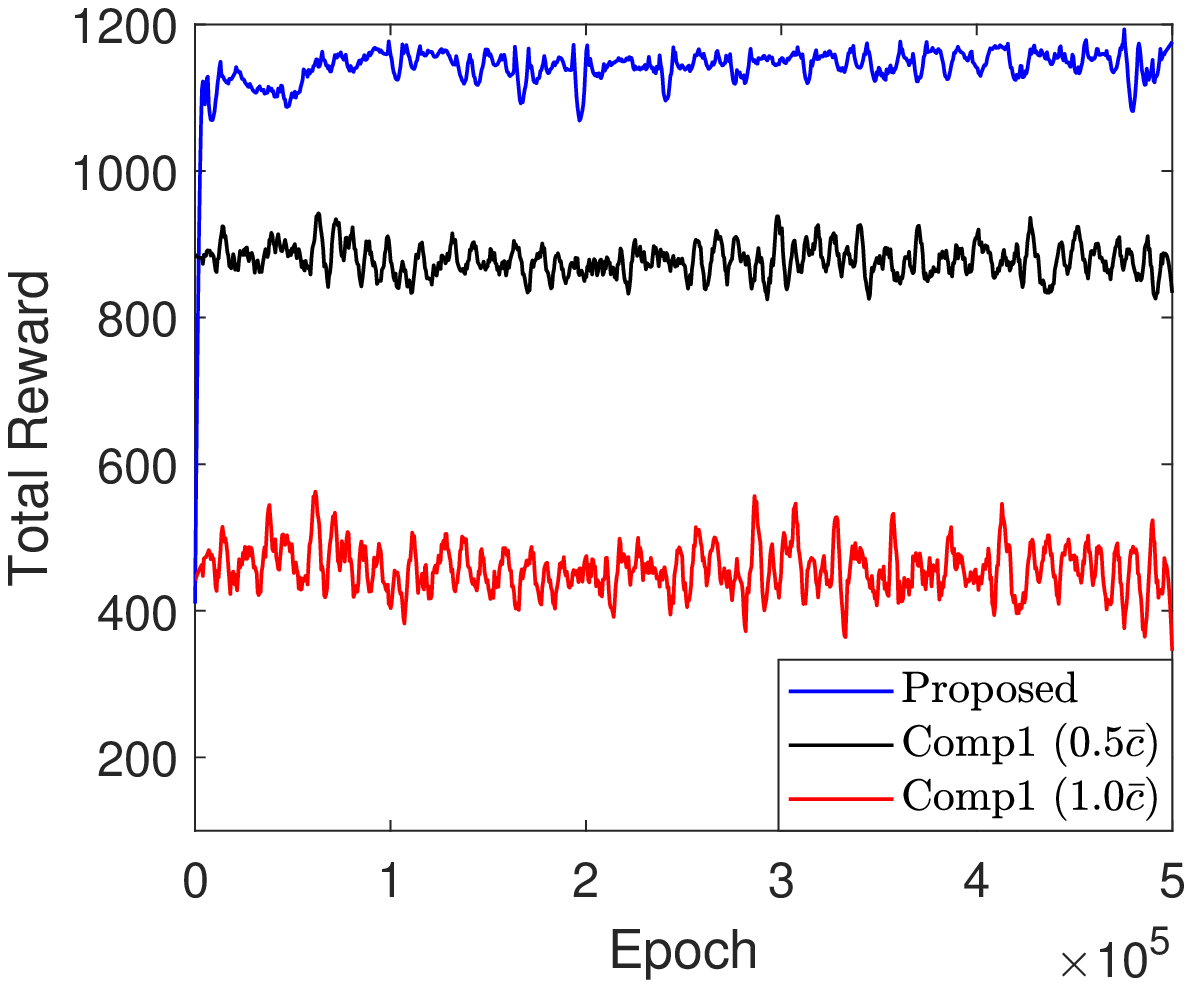}\\
\tabularnewline
\small\centering(e) Packet drops at queues of mBSs. & 
\small\centering(f) Packet drops at buffers of vehicles. &
\small\centering(g) Transmission Efficiency. & 
\small\centering(h) Total reward.\\
\tabularnewline
\end{tabular}
\caption{(a)--(g) show the measurable parameters of the system and (h) shows the total reward in DDPG-based Deep Reinforcement Learning.}
\label{fig:experiment1}
\end{figure*}

We design two comparison techniques to verify the advantages of our proposed pushing scheme and they are explained as follows:
\begin{itemize}
    \item \textit{Random Action (Comp1)}: 
    This comparison scheme randomly chooses $l_j^{(i,q)}(t) \sim \textbf{Unif}[0,\tilde{l}]$ for all $x_j \in \mathcal{X}$, $u_i \in \mathcal{U}_j$, $q\in \mathcal{Q}$, and $t\in\{1,\cdots,T\}$, which means that pushing video chunks from the MBS to mBSs is random without consideration of buffer states and locations of vehicles. 
    This scheme is denoted in figures by `Comp1 ($\tilde{l}$)', and we test two values of $\tilde{l}=0.5 \bar{c}, 1.0 \bar{c}$.
    
    \item \textit{Proactive caching in \cite{TWC2016Qiao} (Comp2)}: 
    The proactive caching scheme in \cite{TWC2016Qiao} is also mobility-aware and designed for video streaming in mmWave vehicular networks with the help of RSUs on the road. 
    Although the considered mmWave vehicular scenario is similar to our model, the scheme in \cite{TWC2016Qiao} does not consider the video quality, makes actions only when vehicles are in transit to the next cell, and is proposed by using dynamic programming. 
    Compared to this comparison work, our proposed model operates in real-time and handles video quality based on the DDPG-based model. 
    Therefore, this comparison scheme is modified to the DDPG-based model in order to operate in real-time with static video quality. 
    The state space and action space of \textit{Comp2} are given as our proposed model, and the reward structure of \textit{Comp2}, which does not consider the video quality is given as follows:
    \begin{align}
    R(s_t,a_t;q^{Comp2}) = &\sum\limits_{x_j \in \mathcal{K}}\sum\limits_{u_i \in x_j } \left[ b_i(t) - \mathcal{F} \right] \\\nonumber
    & -\sum\limits_{x_j \in \mathcal{K}}\left[ \tilde{c_j}^{(i)}(t) -l_j^{(i)}(t) +\bar{c}\right] \label{eq:comp2}
    \end{align}
    This comparison scheme is denoted in figures by `Comp2 ($q^{Comp2}$)', and we test three static qualities of $q^{Comp2} \in \{q_1, q_3, q_5\}$.
\end{itemize}


We evaluated the performances of the proposed pushing scheme and streaming system for a mmWave vehicular network by designing and implementing the following three categories of experiments. 
\begin{enumerate}[label=\arabic*)]
\item \textit{Performance comparison work}:
The performances of our proposed DDPG-based video streaming system are compared with the comparison schemes introduced earlier (i.e., \textit{Comp1} and \textit{Comp2}). 
We evaluate various key QoE metrics of video streaming: backhaul usage, average quality, average quality variations, playback stall rate, and packet drop rate.

\item \textit{Robustness to scalable vehicular network}: 
The proposed adaptive streaming system is applied to vehicular networks of enlarged sizes to verify whether the system is robust to scalability.
We test the three different numbers of mBSs as follows: $\mathcal{N}=10$, $\mathcal{N}=20$, and $\mathcal{N}=40$.

\item \textit{Impact of traffic model on learning phase}: 
Overall, we assume the identical velocity $\bar{v}$ for all vehicles; however, in the practical scenario, velocities of vehicles are time-varying.
Therefore, we test the scenarios with different $\bar{v}$ values by using the learning rate optimized for $\bar{v} = 70$ km/h.

\item \textit{Data-intensive simulations with real-world highway trace data}: 
This data-intensive simulation is conducted with the streaming systems based on real-world highway trace data in order to compare with the FSMC model~\cite{otis}.
\end{enumerate}

\begin{figure}[t]
	\centering
	\includegraphics[width=0.8\columnwidth]{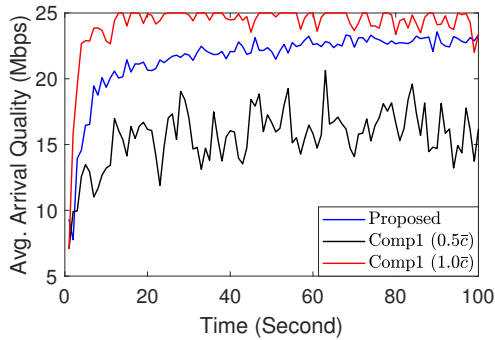}
	\caption{Quality over time in a given episode.}
	\label{fig:qual_time}
\end{figure}

\subsection{Performance Comparisons} 
\label{subsec_V_II}
Figs.~\ref{fig:experiment1} shows how performance metrics and rewards change during the learning phase.
Overall, all of the learning results converge after about 420,000 episodes.
The backhaul usage, the average video quality, average quality variations, playback stall events, packet drops at queues of mBSs, and packet drops at buffers of vehicles, transmission efficiency and total reward are shown in Figs.~\ref{fig:experiment1} (a)--(h), respectively. 
Here, transmission efficiency indicates the ratio of chunks actually delivered to vehicles and chunks pushed from the MBS to mBSs. 
In Fig. \ref{fig:experiment1} (h), we can see that the total reward converges in short training time. 

First of all, the proposed scheme outperforms \textit{Comp1} in terms of all of the performance metrics. 
Note that \textit{Comp1} can provide many chunks with relatively high quality when $\bar{c}$ is large; therefore, the average quality of \textit{Comp1} $(1.0\bar{c})$ is larger than that of \textit{Comp1} $(0.5 \bar{c})$ but still much smaller than that of the proposed scheme. 
Since \textit{Comp1} does not have any intelligence for pushing contents and cannot predict which contents will be desired by vehicles in near future, even when many chunks are pushed from the MBS to mBSs, many of these chunks could not be requested.
This leads inefficient transmission as shown in Fig. \ref{fig:experiment1} (g), meaning that although backhaul usage of \textit{Comp1} is comparable to that of the proposed scheme, its average quality and packet drop rates are all much worse than the proposed scheme. 

Meanwhile, because \textit{Comp2} does not adapt bitrates of the stream; therefore, its quality is fixed and there is no quality variation. 
Also, the structure of the reward function of \textit{Comp2} is completely different from that of the proposed scheme; therefore, the total reward of \textit{Comp2} is not shown. 
Since \textit{Comp2} decides the caching action only when vehicles are in transit to the next mBS coverage, its backhaul usage and packet drop rates are quite smaller than the proposed scheme which makes pushing decisions in every time slot. 
We can see that the proposed scheme can provide higher quality than \textit{Comp2} $(q_1)$ and \textit{Comp2} $(q_3)$ in Fig. \ref{fig:experiment1} (b) while keeping relatively low quality variations. 
On the other hand, \textit{Comp2} $(q_5)$ pushes the highest-quality chunks with small backhaul usage and no quality variation. 
However, packet drop rates of \textit{Comp2} $(q_5)$ become large because it always delivers chunks with $q_5$ having the largest file size. 
The packet drops at both mBS and vehicle sides of \textit{Comp2} $(q_5)$ are slightly larger that those of the proposed scheme, leading that the transmission efficiency of the proposed scheme is better than that of \textit{Comp2} $(q_5)$. 
The more critical point is high playback stall rates for \textit{Comp2}, which is the most important QoE metric of video streaming because users generally prefer a smooth playback even with the quality degradation. 
The average playback stall rates of all \textit{Comp2} schemes are much larger than those of the proposed scheme and \textit{Comp1} when the epoch is small. 
The highly time-varying network dynamics could not allow sufficient training time for the DDPG agent; in this case, the high playback stalls of \textit{Comp2} can be a fatal weakness. 
Also, when the channel condition is not good enough to continuously deliver chunks having large file sizes, \textit{Comp2} runs a risk of frequent playback stalls. 

Note that Figs. \ref{fig:experiment1} (b) and (c) show the average quality and the average quality variations for each episode, not the quality changes over time.
Fig. \ref{fig:qual_time} shows how the average quality of the received chunks changes while vehicles are playing the stream and moving forward.
As shown in Fig. \ref{fig:qual_time}, the proposed DDPG-based scheme provides smoother streaming without abrupt quality variations at the expense of little quality degradation, compared to random action schemes.
These results are consistent with the performances shown in Figs. \ref{fig:experiment1} (b) and (c).

\begin{figure*}[t!]
\centering
\setlength{\tabcolsep}{2pt}
\renewcommand{\arraystretch}{0.2}
\begin{tabular}{p{0.24\linewidth}p{0.001\linewidth}p{0.24\linewidth}p{0.001\linewidth}p{0.24\linewidth}p{0.001\linewidth}p{0.24\linewidth}}
\tabularnewline
\includegraphics[page=1, width=1\linewidth]{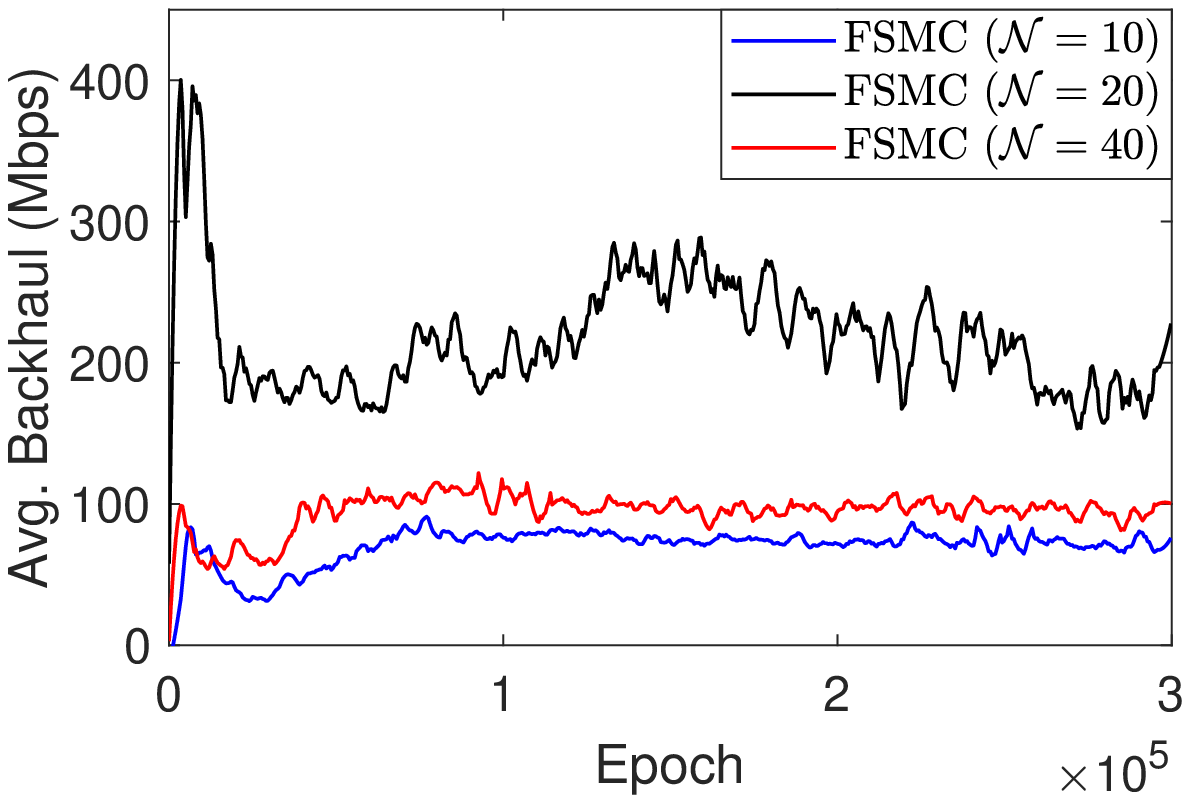}& {} &
\includegraphics[page=1, width=1\linewidth]{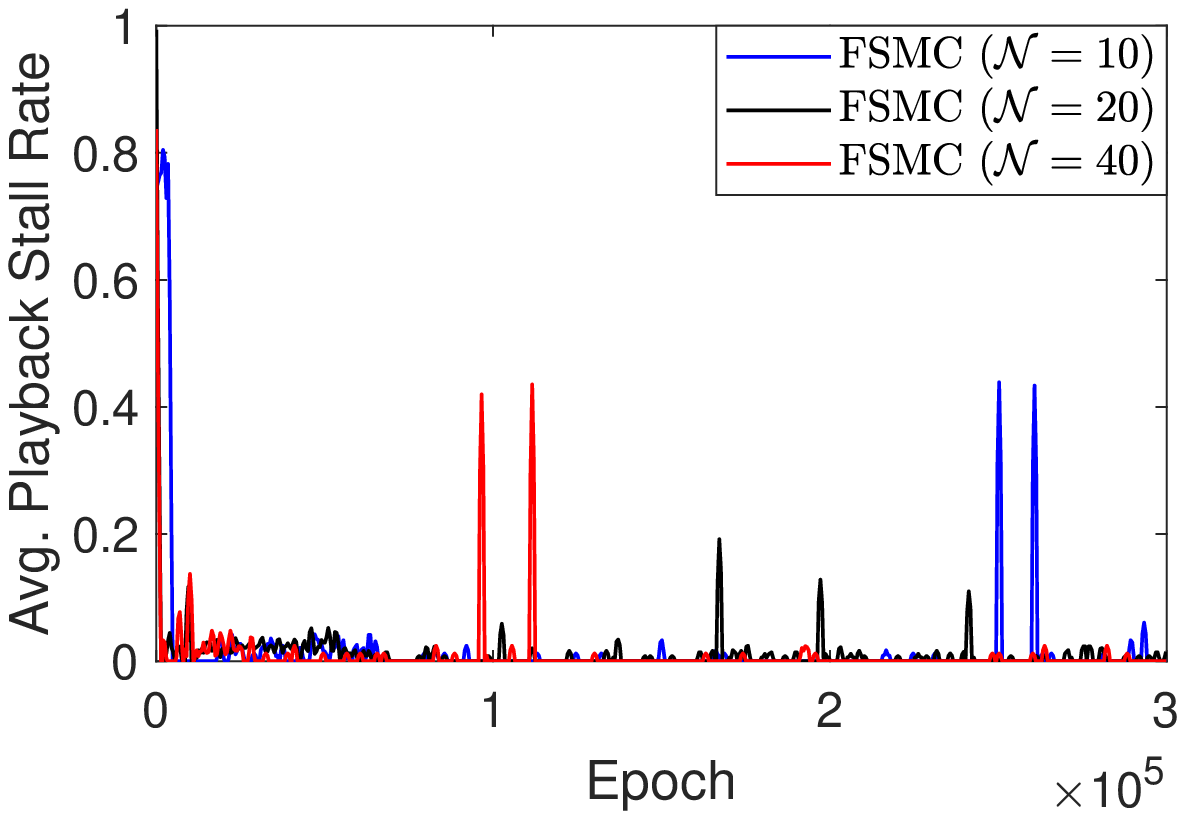}& {} &
\includegraphics[page=1, width=1\linewidth]{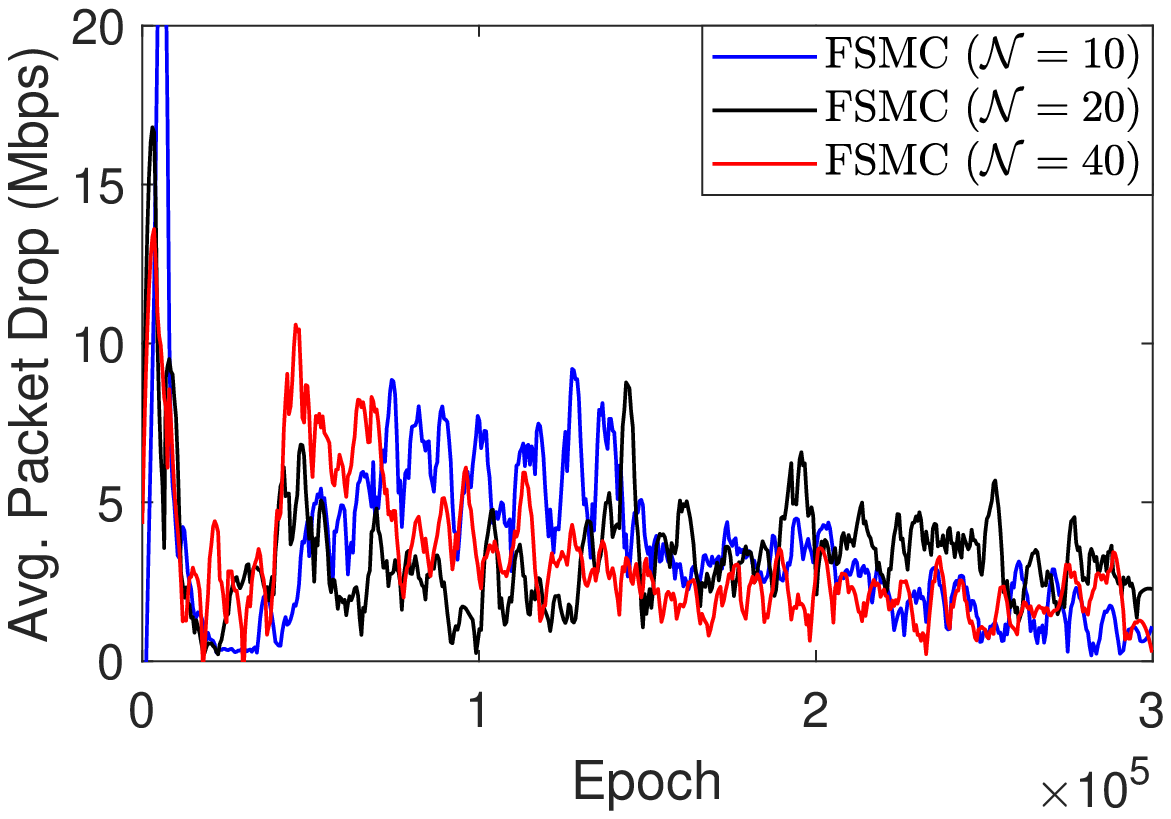} & {} &
\includegraphics[page=1, width=1\linewidth]{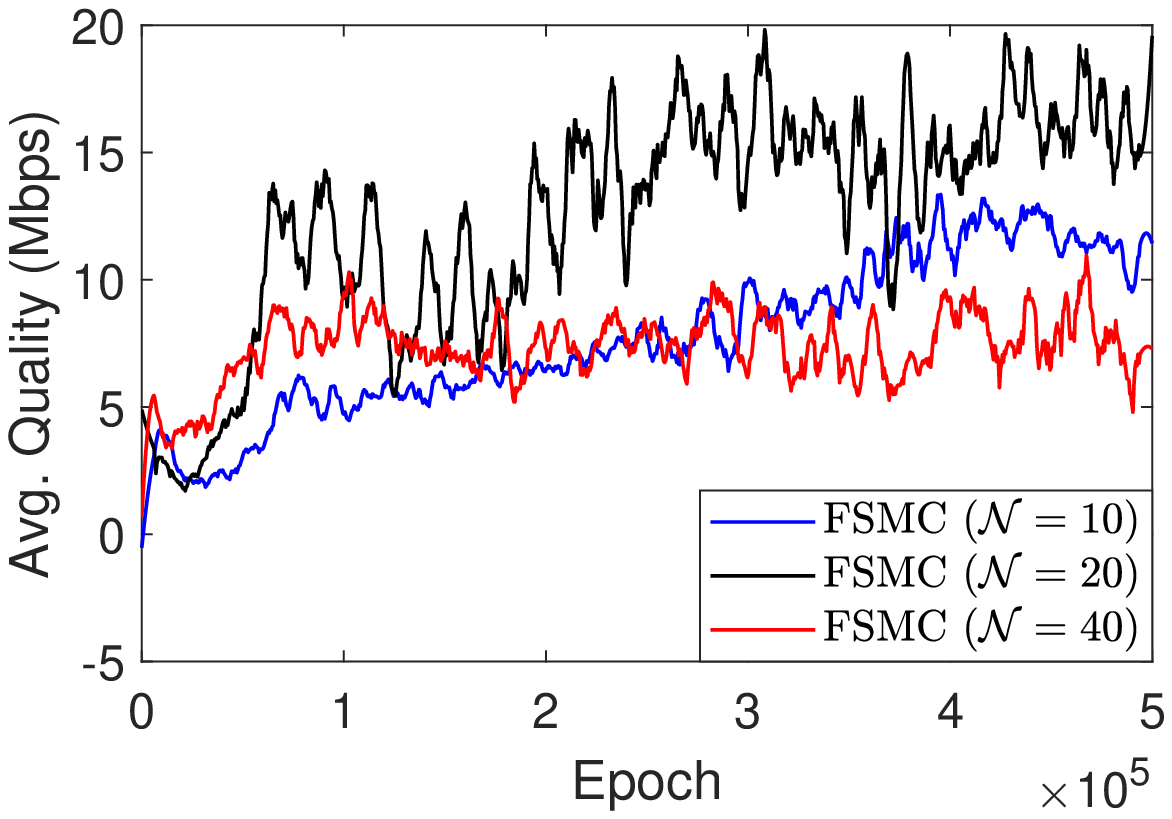}\\
\tabularnewline
\centering(a) Backhaul. & {} &
\centering(b) Playback stall. & {} &
\centering(c) Packet drops at mBS. & {} &
\centering(d) Quality.
\end{tabular}
\caption{Robustness to scalable vehicular networks}
\label{fig:experiment3}
\end{figure*}
\begin{figure*}[t!]
\centering
\setlength{\tabcolsep}{2pt}
\renewcommand{\arraystretch}{0.2}
\begin{tabular}{p{0.24\linewidth}p{0.001\linewidth}p{0.24\linewidth}p{0.001\linewidth}p{0.24\linewidth}p{0.001\linewidth}p{0.24\linewidth}}
\tabularnewline
\includegraphics[page=1, width=1\linewidth]{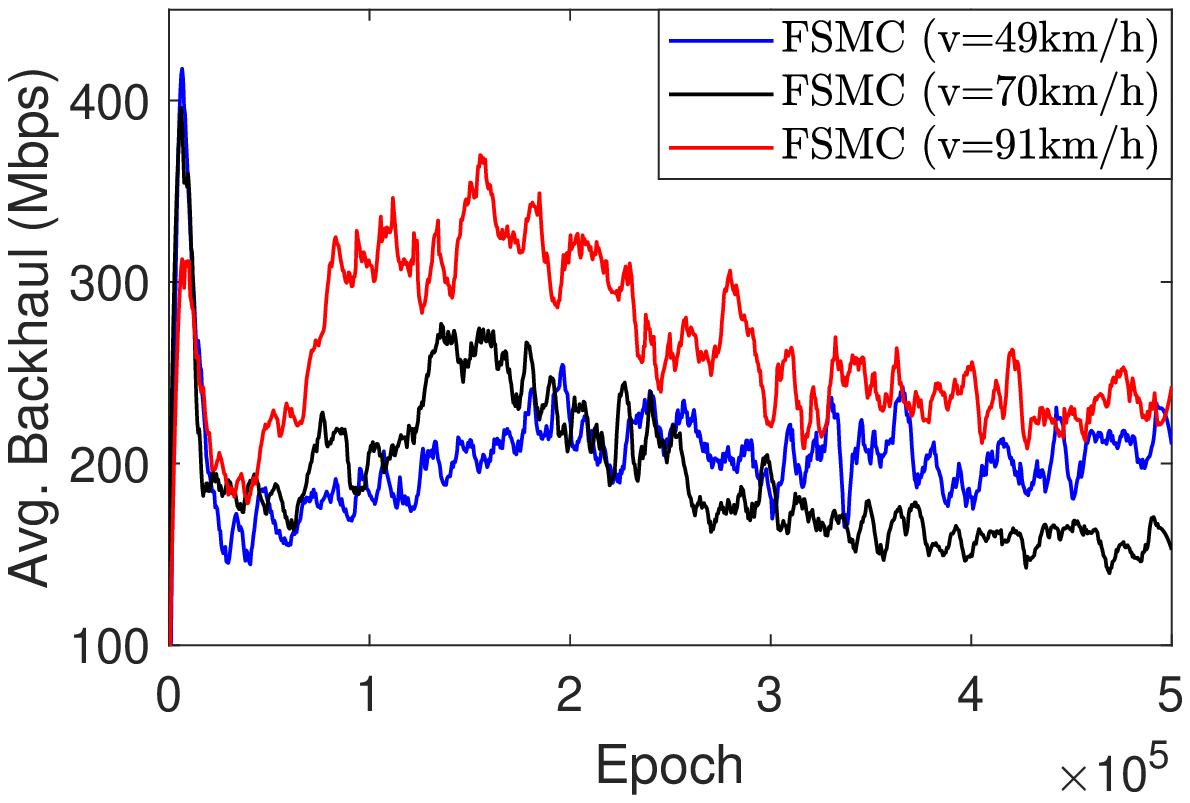} & {} &
\includegraphics[page=1, width=1\linewidth]{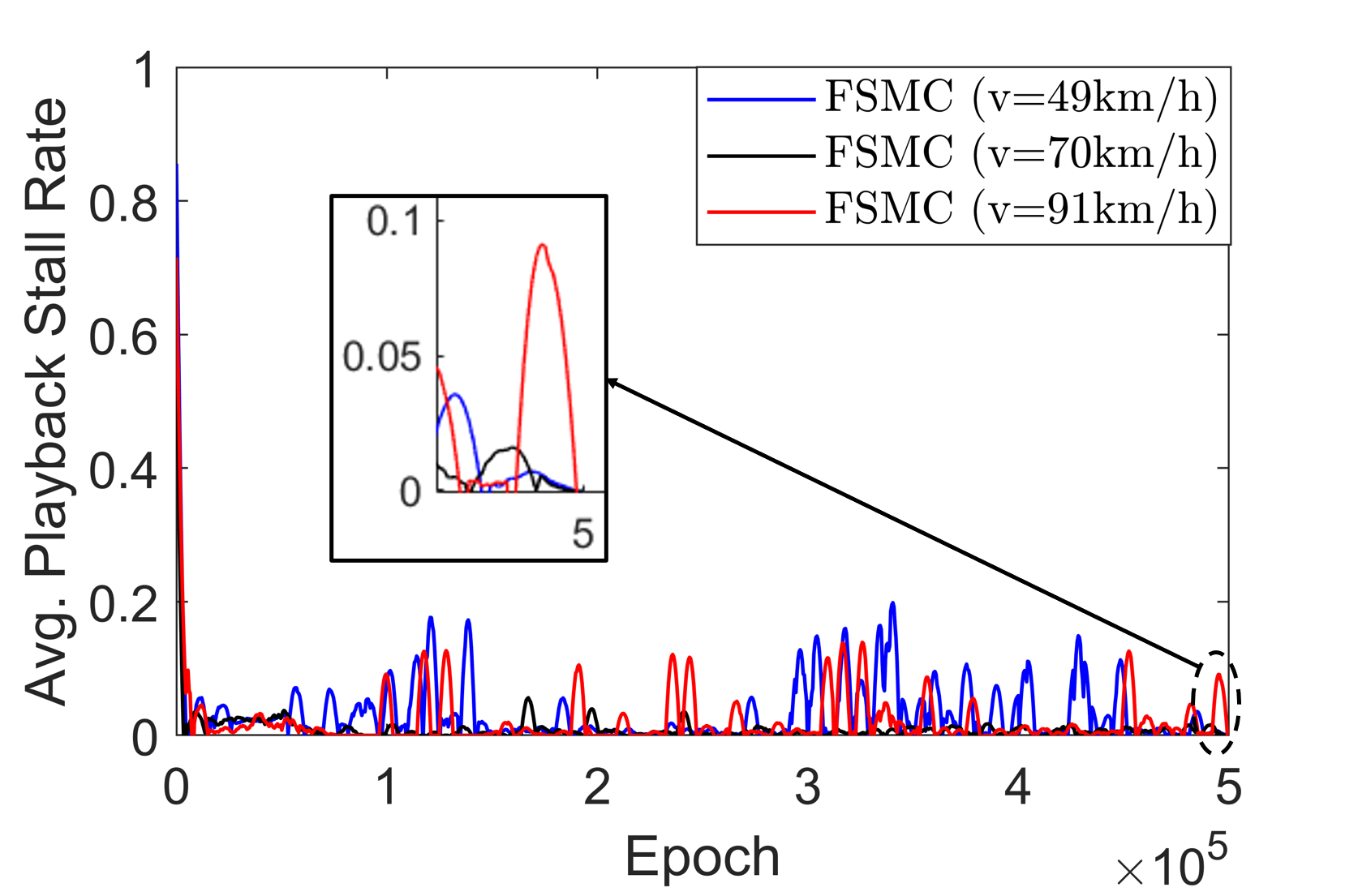} & {} &
\includegraphics[page=1, width=1\linewidth]{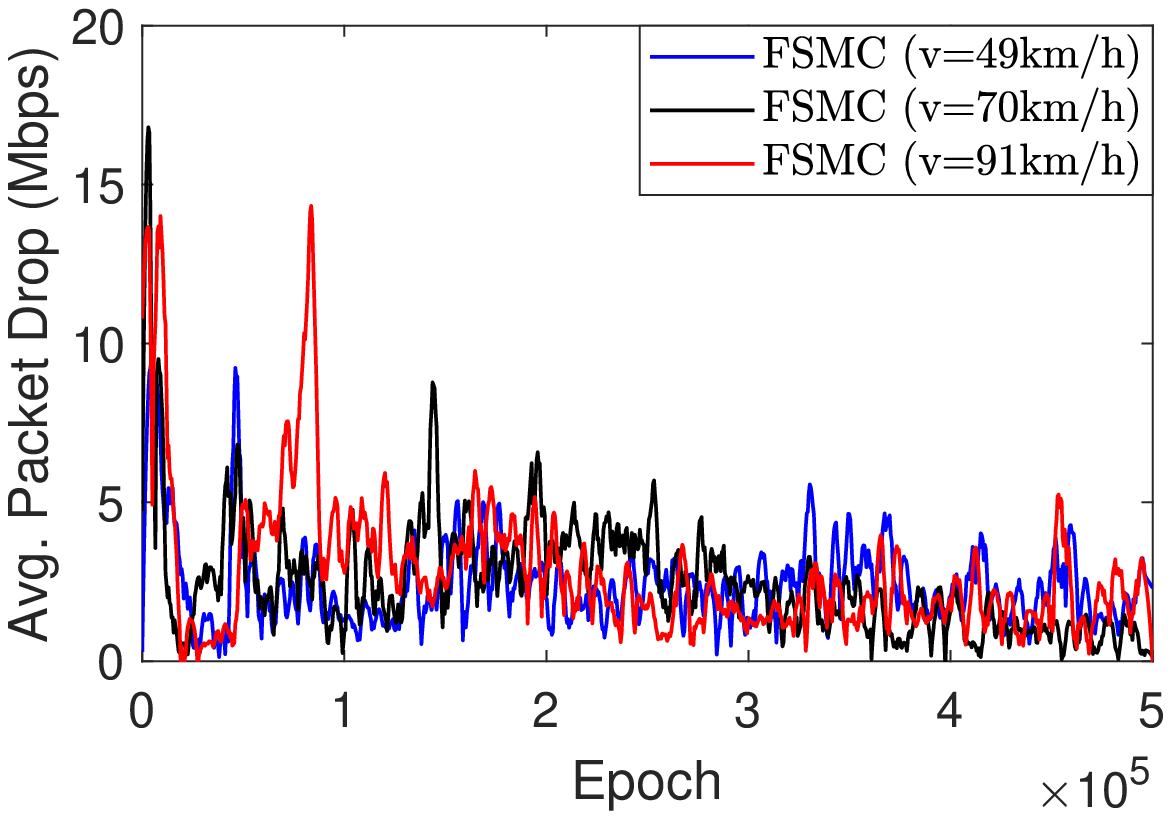} & {} &
\includegraphics[page=1, width=1\linewidth]{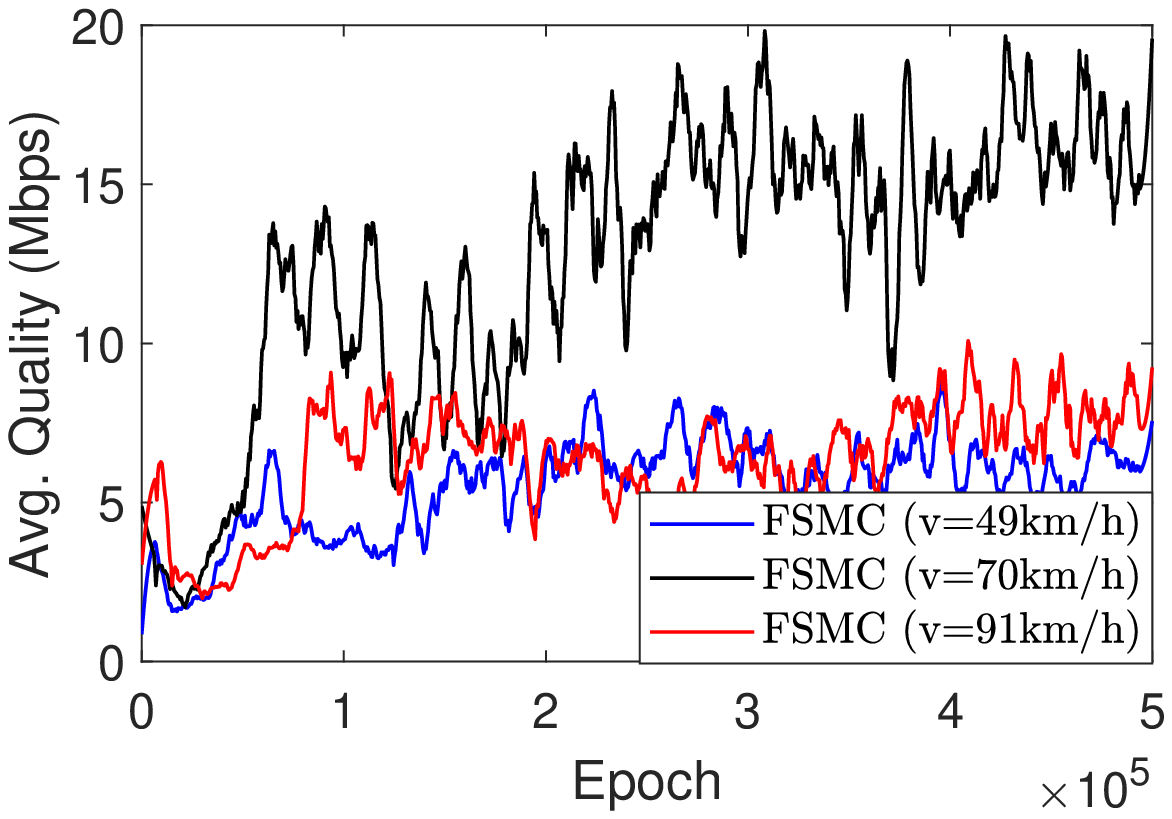}\\
\tabularnewline
\centering(a) Backhaul.  & {} &
\centering(b) Playback stall. & {} &
\centering(c) Packet drops at mBS. & {} &
\centering(d) Quality.
\end{tabular}
\caption{Impact of traffic model on learning phase}
\label{fig:experiment4}
\end{figure*}

\begin{figure*}[t!]
\centering
\setlength{\tabcolsep}{2pt}
\renewcommand{\arraystretch}{0.2}
\begin{tabular}{p{0.24\linewidth}p{0.001\linewidth}p{0.24\linewidth}p{0.001\linewidth}p{0.24\linewidth}p{0.001\linewidth}p{0.24\linewidth}}
\tabularnewline
\includegraphics[page=1, width=1\linewidth]{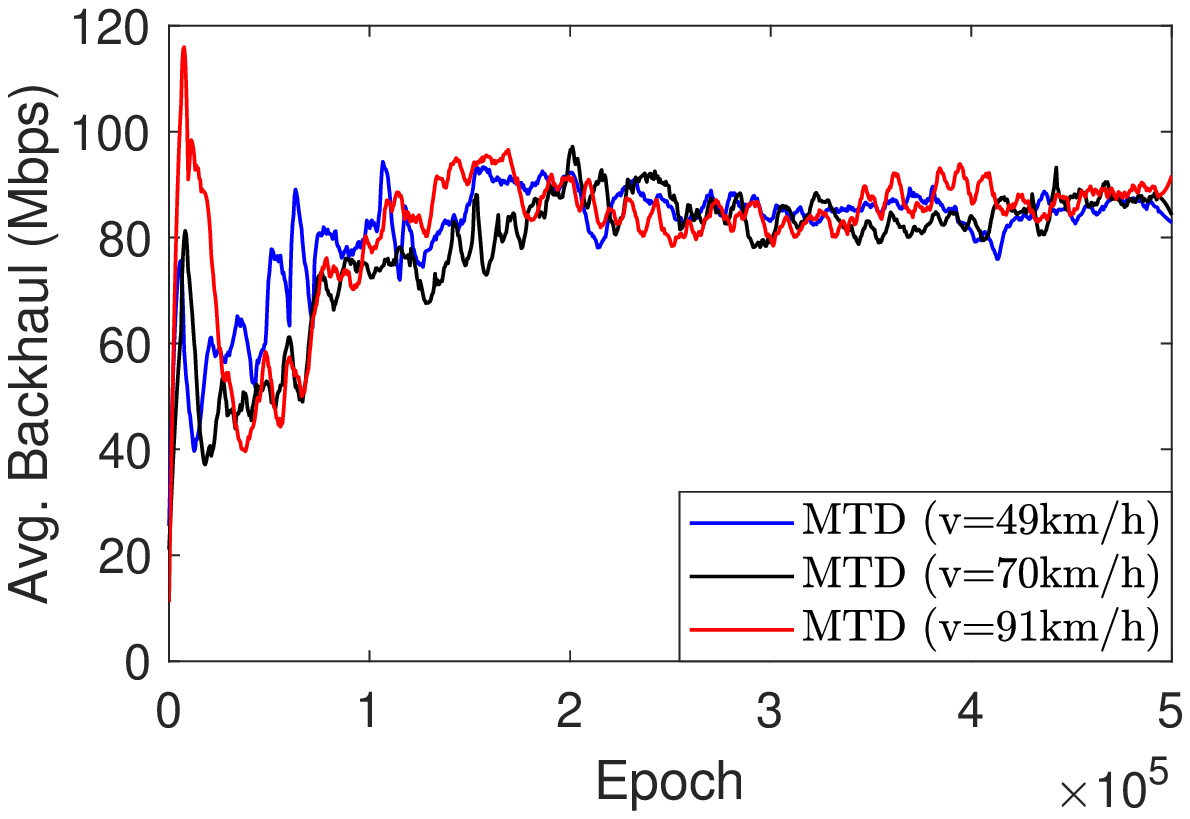} & {} &
\includegraphics[page=1, width=1\linewidth]{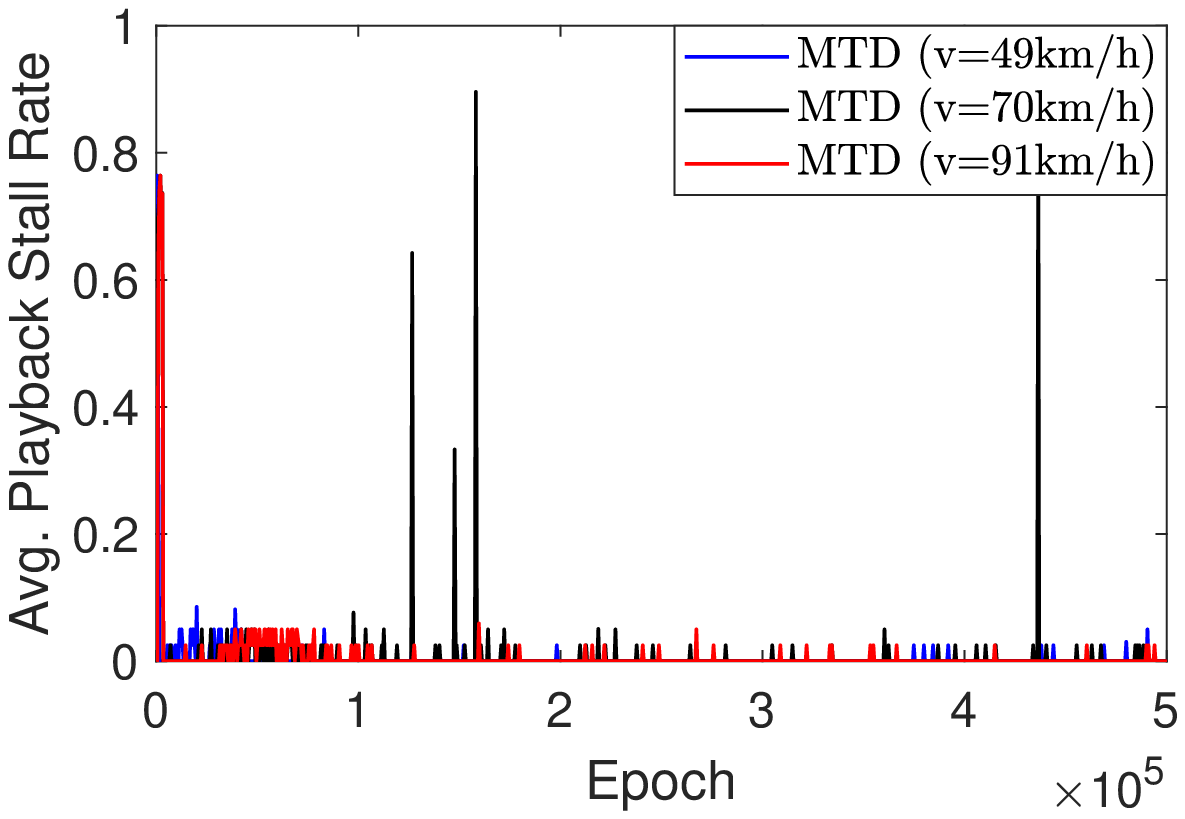} & {} &
\includegraphics[page=1, width=1\linewidth]{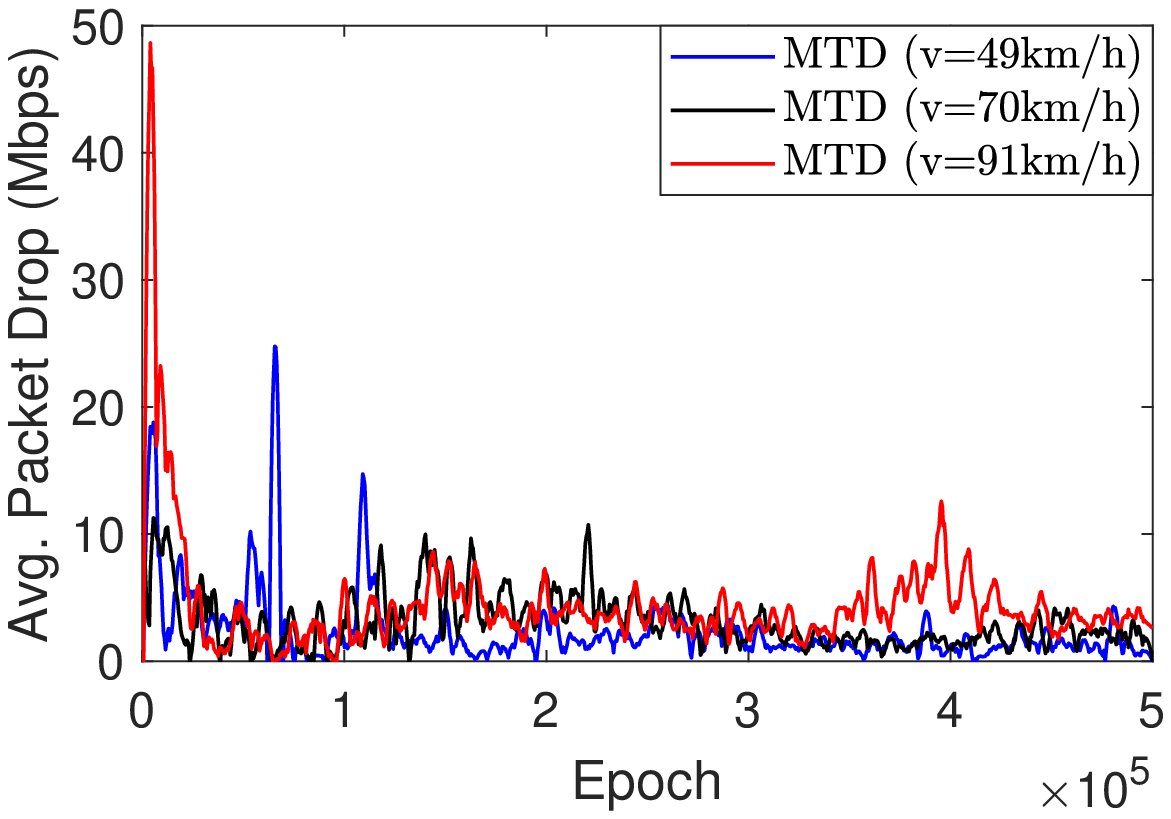} & {} &
\includegraphics[page=1, width=1\linewidth]{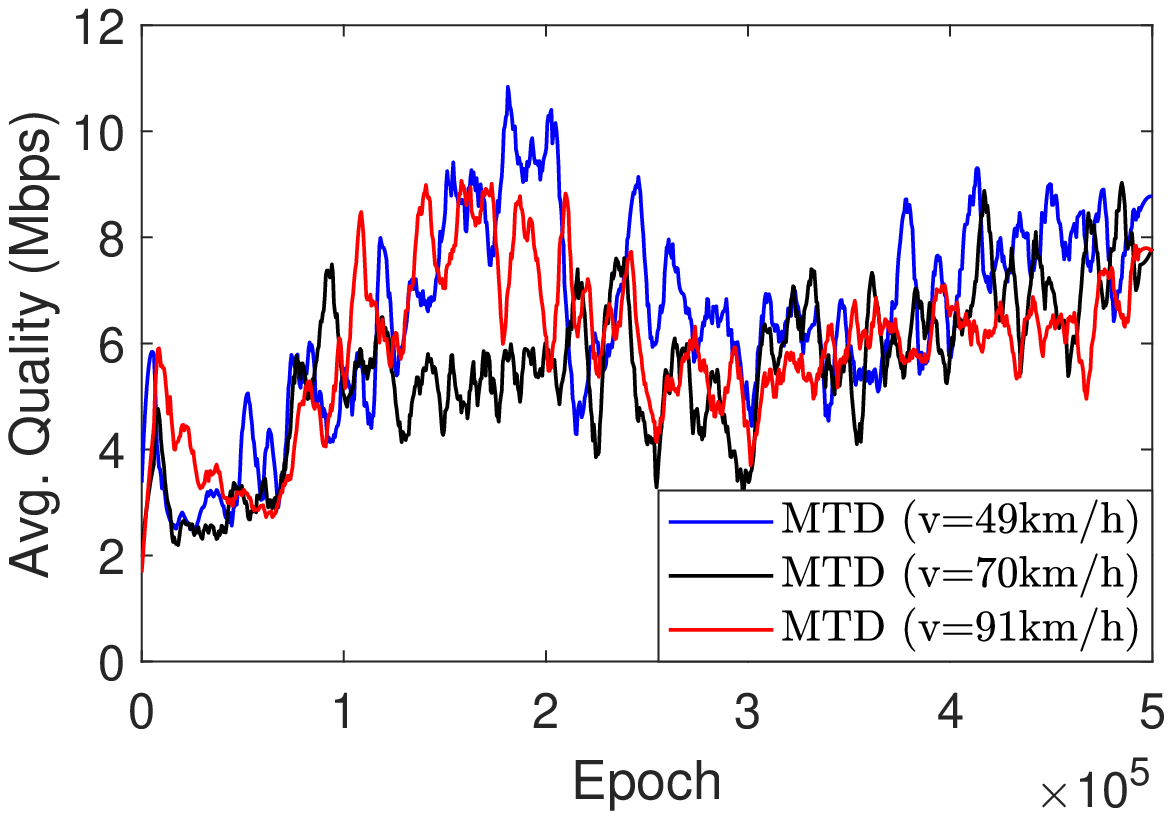}\\
\tabularnewline
\centering(a) Backhaul.  & {} &
\centering(b) Playback stall. & {} &
\centering(c) Packet drops at mBS. & {} &
\centering(d) Quality.
\end{tabular}
\caption{MTD traffic model on learning phase}
\label{fig:experiment5}
\end{figure*}
\subsection{Robustness of Scalable Vehicular Networks} \label{subsec_V_IV}

We considered three different-scale vehicular networks to validate the robustness of the proposed DDPG-based scheme in terms of scalability as follows: $(\mathcal{K}, \mathcal{N}) = (10, 200)$, $(\mathcal{K}, \mathcal{N}) = (20, 200)$, and $(\mathcal{K}, \mathcal{N}) = (40, 200)$. The simulation is conducted with $300,000$ training epochs.
The backhaul usage, playback stalls, packet drops, and the average quality of the above three different networks are shown in Figs.~\ref{fig:experiment3} (a)--(d), respectively.
As $\mathcal{N}$ increases, the MBS has to push more chunks to mBSs for serving more vehicles; therefore, more backhaul bandwidths are required so that backhaul usage of $\mathcal{N}=20$ is larger than that of $\mathcal{N}=10$. 
However, when $\mathcal{N}=40$, backhaul usage decreases again because too many vehicles would make mBSs difficult to provide high-quality chunks to vehicles. 
In order to avoid the playback stall event, the agent can make decisions on pushing low-quality chunks, resulting in decrease in backhaul usage as showin in Fig. \ref{fig:experiment3} (a). 
We can also see in Fig. \ref{fig:experiment3} (d) that the average quality of $\mathcal{N}=40$ is smaller than that of $\mathcal{N}=20$. 
In addition, playback stall and packet drop events are kept very low regardless of $\mathcal{N}$.
Thus, we can conclude that the proposed DDPG agent could control the tradeoff between backhaul usage and the average quality while guaranteeing small playback stall and packet drop rates.


\subsection{Impact of Traffic Model on Learning Phase}\label{subsec_V_V}

This subsection tests different velocities (i.e., $v=49, 70, 91$ km/h) for the proposed DDPG agent which is trained with the assumption of $v = 70$ km/h. 
The backhaul usage, playback stall rate, packet drops, and transmission efficiency with these three different velocities are shown in Figs.~\ref{fig:experiment4} (a)--(d), respectively.
Since the agent is optimized for $v=70$ km/h, the average quality is the highest while keeping low backhaul usage when $v=70$ km/h, compared to the cases of $v = 49, 91$ km/h.
In addition, there is almost no playback stall for $v = 70$ km/h; however, the cases of $v = 49, 91$ km/h show relatively frequent playback stalls. 
Still, the playback stall rates of $v= 49, 91$ km/h remain lower than 0.2 which is still much lower than those of \textit{Comp2} shown in Fig. \ref{fig:experiment1} (d).
Also, packet drop rates of $v=49, 91$ km/h are almost the same as that of $v=70$ km/h.
Therefore, we can say that the proposed adaptive video streaming scheme is capable of serving vehicles with different velocities quite successfully, even using the learning rate optimized for a given velocity value.


\subsection{Simulations with Real-World Highway Trace Data}
\label{subsec_V_VI}
This simulation is conducted with a measured trace data (MTD) model based on Colorado's highway traffic data~\cite{OTIS_RealTraceOfTraffic} as shown in Fig. \ref{fig:experiment5}. 
The proposed DDPG-based scheme provides smooth and high-quality streaming services with small packet rates also in the real traffic data-based MTD model. 
Although there exist some peaks of playback stall rates due to innate unstabiliy of the DDPG algorithm in Fig. \ref{fig:experiment5} (b), it rarely happens so that we can train the agent well with an appropriate epoch number. 
Also, even in the MTD model, the agent trained for $v=70$ km/h still operates well with other velocities (i.e., $v=49, 91$ km/h).
	\section{Conclusion}\label{sec:7}
This paper proposed a DDPG-based adaptive video streaming system for mmWave vehicular networks. 
We present the mBS-assisted streaming scenario where the MBS can push the video chunks to mBSs before vehicles actually request the chunks, and the proposed scheme determines what quality and how many chunks would be provided from the MBS to mBSs. 
For smooth and high-quality video streaming, we model the inventive reward structure for pursuing a variety of QoE metrics of video streaming. 
The simulation results show that the proposed scheme can achieve good performances of backhaul usage, video quality, quality fluctuations, playback stalls, and packet drops at both mBSs and vehicles.
Moreover, our adaptive video streaming method can be properly applied to scalable vehicular networks and scenarios with vehicles of different mobility.

\begin{thebibliography}{1}
		
		\bibitem{Cisco}
		Cisco, "Cisco Visual Networking Index: Global Mobile Data Traffic Forecast Update, 2017--2022 White Paper," 2019. [Online]. Available: https://www.cisco.com/c/en/us/solutions/collateral/service-provider/visual-networking-index-vni/white-paper-c11-738429.html
		
		\bibitem{XuWC2017}
		C. Xu, P. Zhang, S. Jia, M. Wang, and G. M. Muntean, ``Video streaming in content-centric mobile networks: Challenges and solutions," \textit{IEEE Wireless Commun.,} vol. 24, no. 5, pp. 157--165, 2017.
		
		\bibitem{BethTWC2016}
		D. Bethanabhotla, G. Caire, and M. J. Neely, ``WiFlix: Adaptive video streaming in massive MU-MIMO wireless networks," \textit{IEEE Trans. Wireless Commun.,} vol. 15, no. 6, pp. 4088--4103, 2016.
		
		\bibitem{HoTMC2017}
		D. Ho, G. S. Park, and H. Song, ``Game-theoretic scalable offloading for video streaming services over LTE and WiFi networks," \textit{IEEE Trans. Mobi. Comp.,} vol. 17, no. 5, pp. 1090--1104, 2017.
		
		\bibitem{DragoICNC2018}
		M. Drago, T. Azzino, M. Polese, C. Stefanovic, and M. Zorzi, ``Reliable video streaming over mmWave with multi connectivity and network coding," \textit{IEEE ICNC,} Mar. 2018, pp. 508--512.
		
		\bibitem{Qualcomm}
		Qualcomm, "White Paper: Exploring the Potential of mmWave for 5G Mobile Access," 2016. [Online]. Available: https://eu-ems.com/event\_images/Downloads/Qualcomm\%20Whitepaper.pdf
		
		\bibitem{TVT2021Jung} 
		S. Jung, J. Kim, M. Levorato, C. Cordeiro, and J.-H. Kim, ``Infrastructure-assisted on-driving experience sharing for millimeter-wave connected vehicles," \textit{IEEE Transactions on Vehicular Technology}, vol. 70, no. 8, pp. 7307--7321, August 2021.
		
		\bibitem{WC2018Sherman}
		J. Qiao, Y. He, and X. S. Shen, ``Improving video streaming quality in 5G enabled vehicular networks," \textit{IEEE Wireless Commun.,} vol. 25, no. 2, pp. 133--139, May. 2018.
		
		\bibitem{TVT2016Kim} 
		J. Kim, S.-C. Kwon, and G. Choi, ``Performance of video streaming in infrastructure-to-vehicle telematic platforms with 60-GHz radiation and IEEE 802.11ad baseband," \textit{IEEE Trans. Veh. Technol.,} vol.  65, no. 12, pp. 10111--10115, Dec. 2016.
		
		\bibitem{JCN2021Choi}
		M. Choi, M. Shin and J. Kim, ``Dynamic video delivery using deep reinforcement learning for device-to-device underlaid cache-enabled Internet-of-vehicle networks," \textit{Journal of Communications and Networks}, vol. 23, no. 2, pp. 117-128, April 2021.
		
		\bibitem{TON2016Kim}
		J. Kim, G. Caire, and A. F. Molisch, ``Quality-aware streaming and scheduling for device-to-device video delivery," \textit{IEEE/ACM Trans. Netw.,} vol. 24, no. 4, pp. 2319--2331, Aug. 2016.
		
		\bibitem{Sutton_RL}
		R. S. Sutton, and A. G. Barto, \textit{Reinforcement learning: An introduction.} The MIT press, 2018.
		
		\bibitem{OTIS_RealTraceOfTraffic}
		OTIS, "Highway Data Explorer," 2018. [Online]. Available: http://dtdapps.coloradodot.info/otis/HighwayData\#/ui/2/0/criteria/065A
		
		\bibitem{ChoiJSAC2018}
		M. Choi, J. Kim and J. Moon, ``Wireless Video Caching and Dynamic Streaming Under Differentiated Quality Requirements," \textit{IEEE Journal on Selected Areas in Communications}, vol. 36, no. 6, pp. 1245-1257, June 2018.
		
		\bibitem{ChoiTWC2019}
		M. Choi, A. No, M. Ji and J. Kim, ``Markov Decision Policies for Dynamic Video Delivery in Wireless Caching Networks," \textit{IEEE Trans. Wireless Commun.}, vol. 18, no. 12, pp. 5705-5718, Dec. 2019.
		
		\bibitem{ChoiTWC2020}
		M. Choi, A. F. Molisch, and J. Kim, ``Joint Distributed Link Scheduling and Power Allocation for Content Delivery in Wireless Caching Networks," \textit{IEEE Transactions on Wireless Communications}, vol. 19, no. 12, pp. 7810--7824, December 2020.
		
		\bibitem{TVT2017Sun}
		L. Sun, H. Shan, A. Huang, L. Cai, and H. He, ``Channel allocation for adaptive video streaming in vehicular networks," \textit{IEEE Trans. Veh. Technol.,} vol. 66, no. 1, pp. 734--747, Jan. 2017.
		
		
		\bibitem{TM2015Miller}
		K. Miller, D. Bethanabhotla, G. Caire, and A. Wolisz, ``A control-theoretic approach to adaptive video streaming in dense wireless networks," \textit{IEEE Trans. Multimedia,} vol. 17, no. 8, pp. 1309--1322, Jun. 2015.
		
		\bibitem{TB2017Yu}
		L. Yu, T. Tillo, and J. Xiao, ``QoE-driven dynamic adaptive video streaming strategy with future information," \textit{IEEE Trans. Broad.,} vol. 63, no. 3, pp. 523--534, Sep. 2017.
		
		\bibitem{arXivHuang2019}
		T. Huang, C. Zhou, R. X. Zhang, C. Wu, X. Yao, and L. Sun, ``Comyco: Quality-aware adaptive video streaming via imitation learning," \textit{arXiv preprint arXiv:1908.02270}, 2019.
		
		\bibitem{arXivPablo2018}
		P. G. Pereira, A. Schmidt, and T. Herfet, ``Cross-layer effects on training neural algorithms for video streaming," \textit{ACM NOSSDAV}, 2018, pp. 43--48.
		%
		\bibitem{TVT2019Guo}
		Y. Guo, R. Yu, J. An, K. Yang, Y. He, and V. C. Leung, ``Buffer-aware streaming in small scale wireless networks: a deep reinforcement learning approach," \textit{IEEE Trans. Veh. Technol.,} vol. 68, no. 7, pp. 6891--6902, Jul. 2019.
		
		\bibitem{Mobihoc2019Bhattacharyya}
    	R. Bhattacharyya, A. Bura, D. Rengarajan, M. Rumuly, S. Shakkottai, D. Kalathil, Ricky K. P. Mok, and A. Dhamdhere. 2019. QFlow: A Reinforcement Learning Approach to High QoE Video Streaming over Wireless Networks. In \textit{Proceedings of the Twentieth ACM International Symposium on Mobile Ad Hoc Networking and Computing (Mobihoc '19)}. Association for Computing Machinery, New York, NY, USA, 251–260.
		
		\bibitem{arXiv2019Tang}
		K. Tang, N. Kan, J. Zou, X. Fu, M. Hong, and H. Xiong, ``Multiuser video streaming rate adaptation: a physical layer resource-aware deep reinforcement learning approach," \textit{arXiv preprint arXiv:1902.00637}, 2019.
		
		\bibitem{SIGCOMM2017Mao}
		H. Mao, R. Netravali, and M. Alizadeh, ``Neural adaptive video streaming with pensieve," In \textit{ACM SIGCOMM}, Aug. 2017.
		
		\bibitem{TCOMM2019Ye}
    	C. Ye, M. C. Gursoy and S. Velipasalar, ``Power Control for Wireless VBR Video Streaming: From Optimization to Reinforcement Learning," \textit{IEEE Transactions on Communications}, vol. 67, no. 8, pp. 5629-5644, Aug. 2019.
    	
    	\bibitem{TVT2020Guo}
    	Y. Guo, F. R. Yu, J. An, K. Yang, C. Yu and V. C. M. Leung, ``Adaptive Bitrate Streaming in Wireless Networks With Transcoding at Network Edge Using Deep Reinforcement Learning," \textit{IEEE Transactions on Vehicular Technology}, vol. 69, no. 4, pp. 3879-3892, April 2020.
    	
    	\bibitem{IoTJ2021Fu}
    	F. Fu, Y. Kang, Z. Zhang, F. R. Yu and T. Wu, ``Soft Actor–Critic DRL for Live Transcoding and Streaming in Vehicular Fog-Computing-Enabled IoV," \textit{IEEE Internet of Things Journal}, vol. 8, no. 3, pp. 1308-1321, Feb. 2021.
    	
    	\bibitem{TWC2016Qiao}
    	J. Qiao, Y. He and X. S. Shen, ``Proactive Caching for Mobile Video Streaming in Millimeter Wave 5G Networks," \textit{IEEE Transactions on Wireless Communications}, vol. 15, no. 10, pp. 7187-7198, Oct. 2016.
    	
    	\bibitem{CL2019Wu}
    	P. Wu, J. Li, L. Shi, M. Ding, K. Cai and F. Yang, ``Dynamic Content Update for Wireless Edge Caching via Deep Reinforcement Learning," \textit{IEEE Communications Letters}, vol. 23, no. 10, pp. 1773-1777, Oct. 2019.
    	
    	\bibitem{TCCN2020Zhong}
    	C. Zhong, M. C. Gursoy and S. Velipasalar, ``Deep Reinforcement Learning-Based Edge Caching in Wireless Networks," \textit{IEEE Transactions on Cognitive Communications and Networking}, vol. 6, no. 1, pp. 48-61, March 2020.
    	
    	\bibitem{IOTJ2020Wang}
    	X. Wang, C. Wang, X. Li, V. C. M. Leung and T. Taleb, ``Federated Deep Reinforcement Learning for Internet of Things With Decentralized Cooperative Edge Caching," \textit{IEEE Internet of Things Journal}, vol. 7, no. 10, pp. 9441-9455, Oct. 2020.
		
		\bibitem{DDPG}
		T. P. Lillicrap, J. J. Hunt, A. Pritzel, N. Heess, T. Erez, Y. Tassa, and D. Wierstra, ``Continuous control with deep reinforcement learning," \textit{arXiv preprint arXiv:1509.02971}, 2015.
		
		\bibitem{Sutton1999NIPS}
		R. S. Sutton, D. McAllester, S. Singh, and Y. Mansour, ``Policy gradient methods for reinforcement learning with function approximation", \textit{Proceedings of the 12th International  Conference on Neural Information Processing Systems}, pages 1057--1063, 1999.
		
		\bibitem{arXivPreprintQLearning}
		H. Van Hasselt, ``Estimating the maximum expected value: an analysis of (nested) cross validation and the maximum sample average," \textit{arXiv preprint arXiv:1302.7175}, 2013.
		
		\bibitem{ICML2014DPG}
		D. Silver, G. Lever, N. Heess, T. Degris, D. Wierstra, and M. Riedmiller, ``Deterministic policy gradient algorithms," in \textit{Proc. 31st Int. Conf. Mach. Learn. (ICML)}, Beijing, China, Jun. 21--26, 2014, pp.387--395.
		
		\bibitem{ton2011singh}
		S. Singh, R. Mudumbai and U. Madhow, ``Interference Analysis for Highly Directional 60-GHz Mesh Networks: The Case for Rethinking Medium Access Control," \textit{IEEE/ACM Trans. Netw.}, vol. 19, no. 5, pp. 1513--1527, Oct. 2011.
		
		\bibitem{Xavier}
		X. Glorot and Y. Bengio, ``Understanding the Difficulty of Training Deep Feedforward Neural Networks," in \textit{Proc. 13th Int. Conf. Artif. Intell. Stats.}, 2010, pp. 249--256.
		
		\bibitem{ICTC2015Kim}
		J. Kim and E.-S. Ryu, ``Feasibility study of stochastic streaming with 4K
		UHD video traces," \textit{in Proc. Int. Conf. Inf. Commun. Technol. Converg.
		(ICTC)}, Jeju, South Korea, Oct. 2015, pp. 1350–1355.
		
		\bibitem{otis}
		OTIS, ``Highway Data Explorer," 2018. [Online]. Available: http://dtdapps.coloradodot.info/otis/HighwayData
	\end{thebibliography}
\end{document}